%% file: LumFunCov.2.tex
\documentclass[usegraphicx,usenatbib,useAMS]{mn2e} 
\usepackage{amsfonts}
\usepackage{amssymb}
\usepackage{amsmath}
\usepackage{bm}

\input{defs.tex}

\title[How covariant is the galaxy luminosity function?]
{How covariant is the galaxy luminosity function?}

\author[{\it Robert~E.~Smith}]
{
Robert~E.~Smith$^{1,2}$\thanks{res@physik.unizh.ch}
\\
$^1$ Institute for Theoretical Physics, University of Zurich, Zurich CH 8037\\
$^2$ Argelander-Institute for Astronomy, Auf dem H\"ugel 71, D-53121 Bonn, Germany
}

\begin{document}

\maketitle


\begin{abstract}
  We investigate the error properties of certain galaxy luminosity
  function (GLF) estimators. Using a cluster expansion of the density
  field, we show how, for both volume and flux limited samples, the
  GLF estimates are covariant. The covariance matrix can be decomposed
  into three pieces: a diagonal term arising from Poisson noise; a
  sample variance term arising from large-scale structure in the
  survey volume; an {\em occupancy covariance} term arising due to
  galaxies of different luminosities inhabiting the same cluster. To
  evaluate the theory one needs: the mass function and bias of
  clusters, and the conditional luminosity function (CLF). We use a
  semi-analytic model (SAM) galaxy catalogue from the Millennium run
  $N$-body simulation and the CLF of \citet{Yangetal2003} to explore
  these effects. The GLF estimates from the SAM and the CLF
  qualitatively reproduce results from the 2dFGRS.  We also measure
  the luminosity dependence of clustering in the SAM and find
  reasonable agreement with 2dFGRS results for bright
  galaxies. However, for fainter galaxies, $L<L_*$, the SAM
  overpredicts the relative bias by $\sim$10-20\%.  We use the SAM
  data to estimate the errors in the GLF estimates for a volume
  limited survey of volume $V\sim0.13\Gpccube$. We find that different
  luminosity bins are highly correlated: for $L<L_*$ the correlation
  coefficient is ${\bf r}>0.5$. Our theory is in good agreement with
  these measurements. These strong correlations can be attributed to
  sample variance. For a flux-limited survey of similar volume, the
  estimates are only slightly less correlated.  We explore the
  importance of these effects for GLF model parameter estimation. We
  show that neglecting to take into account the bin-to-bin
  covariances, induced by the large-scale structures in the survey,
  can lead to significant systematic errors in best-fit parameters. For
  Schechter function fits, the most strongly affected parameter is the
  characteristic luminosity $L_*$, which can be significantly
  underestimated.
\end{abstract}

\begin{keywords}
Cosmology: large-scale structure of Universe. Galaxies: abundances.
\end{keywords}


\section{Introduction}

The galaxy luminosity function (hereafter GLF) is one of the central
pillars of modern observational cosmology. Commonly denoted $\phi(L)$,
it informs us about the comoving space density of galaxies, per unit
luminosity interval $L$ to $L+dL$. Its central importance originates
through the following: it enables one to quantify the mean space
density of galaxies in a patch of space; it provides a means for
quantifying the evolution over time of the galaxy population in the
Universe; it is one of the main tools for testing models of galaxy
formation; finally it plays a central role in large-scale structure
work, in the construction of mock galaxy catalogues and sample
weighting for clustering estimates.

There is a vast and rich literature on this subject that goes back to
\citet{Hubble1936}, and for a review of developments through to the
mid 90's see the reviews by \citet{Binggelietal1988} and
\citet{StraussWillick1995} and references there in.  Over the past
decade the invention of massive multi-object spectrographs has
revolutionised this area of research and has led to an explosion in
the number of available redshifts with which to estimate the GLF: at
low redshifts there has been the 2dFGRS
\citep{Folkesetal1999short,Coleetal2001short,Norbergetal2002bshort,Crotonetal2005short},
the SDSS \citep{Blantonetal2001short,Blantonetal2003short}, and GAMA
\citep{Lovedayetal2012short} surveys; and at higher redshifts the VVDS
\citep{Ilbertetal2005short}, DEEP2
\citep{Willmeretal2006short,Faberetal2007short}, and the zCOSMOS
\citep{Zuccaetal2009short}.

Our current astrophysical understanding of what shapes the GLF is
evolving rapidly, as our understanding of how galaxies form also
rapidly improves
\citep{KauffmannCharlot1998,Kauffmannetal1999,Coleetal2000,
  Bensonetal2003,Crotonetal2006,Boweretal2010}. This in part owes to
the large spatial volumes that can now be simulated with sufficiently
high enough spatial resolution to follow the growth of dark matter
haloes which may host faint galaxies \citep{Springeletal2005}. One
important insight that has emerged is that there is a quantity more
fundamental than the GLF, and that is the conditional luminosity
function (hereafter CLF) \citep{Yangetal2003,Cooray2006}. This informs
us that the probability of obtaining a galaxy of luminosity $L$, is
conditioned on the mass $M$ of the host halo. This idea is supported
by the results that the GLF is different between dense and void
regions \citep[see for
example][]{Beijersbergenetal2002,Crotonetal2005short}.  This
galaxy--halo connection then provides us with a means for connecting
the estimates of the GLF with the underlying large-scale structures
(LSS).

Whilst the astrophysics that shapes the GLF has been widely studied,
our understanding of the statistical significance of GLF estimates is
far from understood. As we enter an era where the parameterisations of
`good galaxy formation models' are to be compared one needs a more
concrete way of assessing the goodness of fit. Moreover, we would also
like to be able to compare results from different surveys, to make
conclusions about the evolution of the galaxy population. Again, this
requires us to have a more concrete method for interpreting features
and differences. In this paper we aim to provide a theoretical
framework within which one can calculate how large-scale structures
impact not only the shape of the GLF, but also how it shapes the
statistical properties of the errors.  In passing, we note that
\citet{TrentiStiavelli2008} explored how cosmic variance impacts the
GLF parameters for deep high redshift surveys.  We also note that
\citet{Robertson2010} explored a Fisher matrix approach to forecasting
the expected GLFs for future high redshift surveys. He showed that
sample covariance could correlate the galaxy counts in different
magnitude bins. However, as we will show, these authors failed to
capture the full story.  We believe that the formalism presented
herein, goes someway beyond these earlier approaches.

The paper breaks down as follows: in \S\ref{sec:estimators} we present
an overview of some commonly used GLF estimators.  In
\S\ref{sec:LFVolLim} we examine the expectation and covariance of the
GLF estimator for volume limited samples. In \S\ref{sec:LFFluxLim} we
we do the same but for flux limited samples.  In \S\ref{sec:models} we
describe empirical results for the GLF. We also describe the SAM
galaxy catalogues that we use and also the CLF model that we
employ. Here we also explore the luminosity dependence of galaxy
clustering. In \S\ref{sec:results} we present our results for the
error properties of the GLF in volume and flux-limited surveys. In
\S\ref{sec:parameters} we explore the importance of including the full
data covariance matrix for model fitting and parameter estimation.
Finally, in \S\ref{sec:conclusions} we summarise our findings and draw
our conclusions.


\section{Estimating Luminosity Functions}\label{sec:estimators}

\subsection{$\Lambda$CDM paradigm}

Let us begin our theoretical development by following the standard
paradigm for galaxy formation in a $\Lambda$CDM universe: we assert
that galaxies can only form inside dark matter haloes, and that halo
formation, and hence galaxy formation, takes place
hierarchically. Thus, massive galaxies are assembled through the
accretion and merger of smaller ones. Thus, given a dark matter halo
of mass $M$, the detailed theory of galaxy formation will tell us
important information such as, the number, luminosity and types of
galaxies that form inside such haloes. This of course will be a
stochastic process and the exact number will vary between haloes.


\subsection{Overview of estimators}

One of the most basic observational tools for testing our
understanding of galaxy formation models is through the GLF.  Over the
years there have been many approaches to constructing estimators for
the GLF. The simplest is to compute:
\be {\rm E1:}
\ \ \ \widehat{\phi}_1(L_{\mu})=\frac{N^{\g}(L_{\mu})}{\Vs\Delta L_{\mu}}
\ ,\label{eq:E1} \ee
where $N^{\g}(L_{\mu})$ is the number of galaxies of luminosity
$L_{\mu}$ in the bin $\Delta L_{\mu}$, and $\Vs$ is the total sampled
survey volume.

For flux limited surveys this proves to be a biased estimator, since
for faint galaxies the volume out to which one may observe these
objects is significantly smaller than for the case of bright
galaxies. This can be corrected for by adopting the $V^{\rm max}$
estimator of \citet{Schmidt1968}:
\be {\rm E2:} \ \ \ \widehat{\phi}_2(L_{\mu})=\frac{1}{\Delta
  L_{\mu}}\sum_{\mu=1}^{N^{\rm g}(L_{\mu})}\frac{1}{\Vmu}\ , \label{eq:E2}\ee
where $\Vmu\equiv V^{\rm max}(L_{\mu})$ is the maximum volume that a
galaxy with a luminosity $L_{\mu}$ could have been found in, given the
flux limit of the survey $m_{\rm lim}$ (for further details see
\S\ref{sec:LFFluxLim}). For a discussion of estimators E1 and E2 see
\citet{Felten1976} and references therein.

It was noted that for shallow and narrow surveys estimators E1 and E2
would be `biased' by the presence of large-scale over/underdense
regions. Subsequently, a further set of estimators were developed to
try and remove this so called bias
\citep{Turner1979,Sandageetal1979,Kirshneretal1979,Efstathiouetal1988}. At
the heart of these approaches is the assumption that the joint
probability of obtaining a galaxy with luminosity $L_{\mu}$ in interval
$\Delta L_{\mu}$, and spatial position in the volume element $\dx$, is
the product of two independent probability density functions (PDF):
\be p(L_{\mu},\bx) \dL_{\mu} \dx = p(L_{\mu}) p(\bx) \dL_{\mu}\dx \ , \label{eq:ass} \ee
where the 1-point luminosity PDF is
\be 
p(L)=\frac{\phi(L)}{\Phi(L_{\rm min})} \ ; \ \hspace{0.2cm}
\Phi(L)\equiv \int_{L}^{\infty}\dL' \phi(L') \ .
\ee
where $L_{\rm min}$ is the lowest luminosity galaxy detectable in the
sample volume, given selection criteria.  If $\bx$ is the location of
a random point then the probability of finding a galaxy in a cell of
volume $\delta V$ is given by:
\be P(\bx)=p(\bx)\dx=N \delta V/\Vs = \bar{n} \delta V\ .\ee
However, if one pre-selects a cluster region centred on $\bx_c$, then
the probability is enhanced $P(\bx|\bx_{c})=\nbar\delta V [1+\xi_{\rm
  gc}(r)]$, where $r=|\bx-\bx_c|$ and $\xi_{\rm gc}(r)$ is the
cross-correlation function between the cluster centre and galaxies in
the cluster \citep{Peebles1980}.  Then, for example for estimator E1,
the luminosity function estimate would be:
\be {\rm E1}: \widehat{\phi}(L_{\mu}) = \frac{N^{\g}(L_{\mu})}{\Vs\Delta
  L_{\mu}} = \left<\phi(L_{\mu})\right>\left[1+\sigma^2\right] \ ,\ee
where 
\be \sigma^2\equiv\sum_{i=1}^{N^{\rm g}(L_{\mu})}\xi_{\rm
  gc}(r_i)/N^{\rm g}(L_{\mu}) \ . \ee

\citet{Turner1979} saw that, under the assumption of \Eqn{eq:ass}, if
one constructed the following quantity, then the environmental
dependence of the counts would drop out:
\ba {\rm E3}: \frac{dN^{\g}(L_{\mu})}{N^{\g}[>L_{\mu},\chi\le\chi^{\rm
      max}(L_{\mu})]} & = & \frac{\phi(L_{\mu})\dL_{\mu} \times
  p(\bx)\Vs} {\int_{L_{\mu}}^{\infty}\dL' \phi(L') \times p(\bx)\Vs }
\nn \\ & = & \frac{\phi(L_{\mu})\dL_{\mu}
}{\int_{L_{\mu}}^{\infty}\dL' \phi(L') }\ , \ea
where $N^{\g}[>L_{\mu},\chi\le\chi^{\rm max}(L_{\mu})]$ denotes the
total number of galaxies brighter than $L_{\mu}$ with distance less
than $\chi^{\rm max}(L_{\mu})$.

Unfortunately, the estimator E3 is also biased -- the real world is
more complicated \citep[see][for additional discussion of
this]{Cole2011}. The bias can be attributed to the fact that
$p(L,\bx)$ is not separable: bright/faint galaxies tend to inhabit
high/low density environments \citep{Norbergetal2002ashort}. To
illustrate how this bias operates, let us consider the following toy
example.  Suppose our survey consists of two clusters at the same
distance from the observer, and let cluster one contain galaxies of
luminosity $L_1$ and be of mass $M_1$, and let cluster two contain
galaxies of $L_2>L_1$ and be of mass $M_2>M_1$. Then since higher mass
dark matter haloes have more extended profiles and also are more
biased with respect to the underlying dark matter than lower mass
haloes, then we have: $\xi_{\rm gc}(r|L_2)>\xi_{\rm gc}(r|L_1)$. On
construction of Turner's estimator we find:
\ba
& & \hspace{-0.5cm}\frac{dN(L_1)}{N[>L_1,\chi\le\chi^{\rm max}(L)]} \nn \\
& & \hspace{0cm}=\frac{\Vs\Delta L\left<\phi(L_1)\right>\left[1+\sigma_1^2(L_1)\right]}
{\Vs\Delta L \left\{\left<\phi(L_1)\right> \left[1+\sigma^2_1(L_1)\right]+  \left<\phi(L_2)\right> 
\left[1+\sigma^2_2(L_2)\right]\right\}} 
\  \nn \\
& & \hspace{0cm}=
\left\{1+
\frac{\left<\phi(L_2)\right>}{\left<\phi(L_1)\right>} 
\frac{\left[1+\sigma^2_2(L_2)\right]}{\left[1+\sigma^2_1(L_1)\right]}\right\}^{-1} \ ,
\ea
where in the above we have defined
\be \sigma^2_j(L)\equiv
\frac{1}{N(L)}\sum_{i=1}^{N(L)}\xi_{gc}(\bx_i-\bx_{c,j}|L)\ .\ee
%
%
%
Thus we see that, in this toy-model case, the estimator is biased low
for the lower luminosity galaxies. 

\vspace{0.1cm}

In fact as we will show in the following sections the bias associated
with estimators E1 and E2 approaches zero, provided that the sample
volume is sufficiently large. Whereas for estimator E3 one can see
that owing to the fact that $\xi_{\rm gc}(r|L_2)\ne\xi_{\rm
  gc}(r|L_1)$, the estimator is biased.  We shall reserve a more
detailed study of Turner's estimator and the bias induced by
neglecting density-luminosity correlations for future study.


\section{Volume limited galaxy samples}\label{sec:LFVolLim}

Let us consider the simplest estimator E1, which one may apply to
volume limited surveys. We are interested in computing the expectation
and covariance. 


\subsection{Expectation of estimator}

Consider some large cubical patch of the Universe, of volume $\Vs$,
and containing $N^{\rm c}$ clusters that possess some distribution of
masses.  Let us subdivide this set of clusters into a set of $N_m$
mass bins, and where the $\alpha$th mass bin contains $N_{\alpha}^{\rm
  c}$ clusters.  We shall denote the number of galaxies with
luminosities between $L_{\mu}-\Delta L_{\mu}/2$ and $L_{\mu}+\Delta
L_{\mu}/2$, that are hosted by the $i$th halo of the $\alpha$th mass
bin, by $N^{\rm g}_{i,\alpha,\mu}$.

With the above definitions, the GLF estimator E1 for volume limited
samples can be written:
\be 
{\rm E1:}\ \  \widehat{\phi}(L_{\mu}) = 
\frac{1}{\Vs\Delta L_{\mu}}
\sum_{\alpha=1}^{N_M}
\sum_{i=1}^{N^{\rm c}_{\alpha}} N^{\rm g}_{i,\alpha,\mu} \label{eq:E1B} \ .
\ee
We now wish to compute the expectation of this estimator. We shall
write this as,
\be
\left<\widehat{\phi}(L_{\mu})\right> 
=  \frac{1}{\Vs\Delta L_{\mu}} \sum_{\alpha=1}^{N_M}
\left<  \sum_{i=1}^{N^{\rm c}_{\alpha}} 
N^{\rm g}_{i,\alpha,\mu} \right>_{g,P,s} \label{eq:ExpecE1} \ ,
\ee
where in the above $\left<\dots\right>_{g,P,s}$ represents an
averaging over the ensemble: the subscript $g$ denotes an averaging
over the sampling distribution for placing galaxies into haloes; the
subscript $P$ denotes an averaging over sampling clusters into the
given realization of the density field; and the subscript $s$ denotes
an averaging over the density fluctuations within the volume.

We shall assume that the number of galaxies occupying a given dark
matter halo is a Poisson process:
\be 
P(N^{\rm g}_{i,\alpha,\mu}|\lambda_{\alpha,\mu})= 
\frac{\lambda^{N^{\rm g}_{i,\alpha,\mu}}\exp[-\lambda]}{N^{\rm g}_{i,\alpha,\mu}!} \ .
\label{eq:defPoisson}
\ee
where $\lambda\equiv N^{\rm g}(M_{\alpha},L_{\mu})$ is the expected
number of galaxies in the $L_{\mu}$ luminosity bin, and for a halo of
mass $M_{\alpha}$. Actually, the above sampling distribution is not of
great concern, but what will be of importance will be the independence
of the distributions, i.e. the number of galaxies occupying a given
cluster depends only on the physical properties of that cluster.

One immediate consequence of this is that we may compute the average over 
the galaxy population separately, and hence write E1 as:
\ba
\left<\widehat{\phi}(L_{\mu})\right> 
& = &  \frac{1}{\Vs\Delta L_{\mu}} \sum_{\alpha=1}^{N_M} 
\left< \sum_{i=1}^{N^{\rm c}_{\alpha}} 
\left<N^{\rm g}_{i,\alpha,\mu}\right>_g \right>_{P,s} \nn \\
& = &  \frac{1}{\Vs\Delta L_{\mu}} 
\sum_{\alpha=1}^{N_M} N^{\rm g}(M_\alpha,L_\mu)
\left< N^{\rm c}_{\alpha}  \right>_{P,s} \label{eq:ExpecE1_2} \ ,
\ea
where in the last line we identified $N^{\rm
  g}(M_\alpha,L_\mu)\equiv\left<N^{\rm g}_{i,\alpha,\mu}\right>_g$,
which tells us the expected number of galaxies with luminosity in the
interval $[L_{\mu}-\Delta L_{\mu}/2,L_{\mu}+\Delta L_{\mu}/2]$
that occupies a cluster of mass $M$.

In order to proceed further, we need to compute the expected number of
clusters in the $\alpha$th mass bin, $\left< N^{\rm c}_{\alpha}
\right>_{P,s}$. This may be done following the procedure described in
\citet{SmithMarian2011} (summarised in Appendix \ref{app:cluster} for
convenience). Following this procedure gives:
\be
\left< N^{\rm c}_{\alpha}\right>_{P,s} =  \Vs \overline{n}_{\alpha} \ , 
\ee
where the number density of clusters in the $\alpha$th mass bin is
\be 
\overline{n}_{\alpha} \equiv \int_{M_\alpha-\Delta M_{\alpha}/2}^{M_\alpha+\Delta M_{\alpha}/2} dM n(M) \ ,
\label{eq:nbarh}
\ee 
and where $n(M)dM$ is the abundance of dark matter haloes in the mass
interval $[M-dM/2,M+dM/2]$.  On inserting this expression into
\Eqn{eq:ExpecE1_2} we find:
\be 
\left<\widehat{\phi}(L_{\mu})\right> =  \frac{1}{\Vs\Delta
  L_{\mu}} \sum_{\alpha=1}^{N_M} 
N^g(M_{\alpha},L_{\mu}) \Vs \overline{n}_{\alpha}  \ . \label{eq:temp}
\ee
On taking the limit of small mass bins and assuming that the mass
function varies slowly across the bins, then from the mean value
theorem, we have
\be \overline{n}_{\alpha} \approx n(M_{\alpha}) \Delta M_{\alpha}\ . \ee
and we may convert \Eqn{eq:temp} to an integral. Finally, on using the
CLF model of \citet{Yangetal2003}, for which 
$\Phi(L_\mu|M_{\alpha})\equiv N^g(M_{\alpha},L_{\mu})/\Delta L_{\mu}$,
then we have:
\be \left<\widehat{\phi}(L_{\mu})\right> = 
\int dM n(M) \Phi(L_{\mu}|M) \label{eq:LFfromCLF} \ .\ee
Thus for volume limited samples, estimator E1 is unbiased.


\subsection{Estimator covariance}

Let us compute the covariance matrix that we would expect for
estimator E1. The covariance matrix is defined to be,
\be {\mathcal C}_{\mu\nu} \equiv
\left<\widehat{\phi}_\mu\widehat{\phi}_\nu\right> -
\left<\widehat{\phi}_\mu\right> \left<\widehat{\phi}_\nu\right>
\ , \label{eq:LFcov} \ee
where from now on we make use of the compact notation
$\phi_{\mu}\equiv\phi(L_{\mu})$.  Focusing on the first term on the
right-hand-side, and on inserting \Eqn{eq:E1B}, we find
\ba 
\left<\widehat{\phi}_\mu\widehat{\phi}_\nu\right> 
& = & \frac{1}{\Delta L_{\mu}\Delta L_{\nu}\Vs^2}\sum_{\alpha=1}^{N_M}
\sum_{\beta=1}^{N_M} \nn \\
& & \times \left<\sum_{i=0}^{N^{\rm c}_{\alpha}}\sum_{j=0}^{N^{\rm c}_{\beta}}
\left< N^g_{i,\alpha,\mu}N^g_{j,\beta,\nu}\right>_g\right>_{P,s}\ , \label{eq:part0}
\ea
where again we have used the fact that the average over the galaxies
can be separated from the cluster sample.  Considering the contents of
the inner bracket, we see that this may be rewritten as
\ba 
\left< N^g_{i,\alpha,\mu}N^g_{j,\beta,\nu}\right>_g & = &
\tilde{\epsilon}_{ij} \tilde{\epsilon}_{\alpha\beta} \tilde{\epsilon}_{\mu\nu} 
\left<N^g_{i,\alpha,\mu}\right>_g
\left<N^g_{j,\beta,\nu}\right>_g \nn \\
& + & 
\delta^K_{ij} \tilde{\epsilon}_{\alpha\beta} \tilde{\epsilon}_{\mu\nu} 
\left<N^g_{i,\alpha,\mu}\right>_g
\left<N^g_{j,\beta,\nu}\right>_g \nn \\
& + & 
\dots + {(\rm 5\, terms)}
\nn \\
& + & 
\delta^K_{ij}\delta^K_{\alpha\beta} \delta^K_{\mu\nu}  
\left<(N^g_{i,\alpha,\mu})^2\right>_g 
\ \label{eq:part1} ,
\ea
where in the above we have made use of a modified Levi-Cevita symbol
$\tilde{\epsilon}_{ij}=1$ if $i\ne j$ and 0 otherwise, and we have
used the independence of the sampling distributions to separate the
expectations of the products. Consider the final term in the above
expression, on using \Eqn{eq:defPoisson}, we see that this piece can
be rewritten as,
\ba 
\left<(N^g_{i,\alpha,\mu})^2\right>_g & = & \left<N^g_{i,\alpha,\mu}\right>_g^2+
\left<N^g_{i,\alpha,\mu}\right>_g \nn \\
& = & N^g(M_{\alpha},L_{\mu})\left[1+N^g(M_{\alpha},L_{\mu})\right]\ . 
\ea
On inserting this back into \Eqn{eq:part1}, we may resum all terms and
find that the expression simplifies to be,
\ba 
\left< N^g_{i,\alpha,\mu}N^g_{j,\beta,\nu}\right>_g
& = & 
N^g(M_\alpha,L_{\mu})N^g(M_{\beta},L_{\nu}) \nn \\
& + & N^g(M_\alpha,L_{\mu})\delta^{K}_{i,j}\delta^{K}_{\alpha,\beta}\delta^{K}_{\mu,\nu}\ .
\label{eq:galcov}
\ea
If we now return to \Eqn{eq:part0}, then on using the above relation,
we find:
\ba 
\left<\widehat{\phi}_\mu\widehat{\phi}_\nu\right> 
& =  & \frac{1}{\Delta L_{\mu}\Delta L_{\nu}\Vs^2}
\sum_{\alpha=1}^{N_M} \sum_{\beta=1}^{N_M} \nn \\
& & \times \ 
\left[ 
\left<N^{\rm c}_{\alpha}N^{\rm c}_{\beta}\right>_{P,s} 
N^g(M_\alpha,L_{\mu})N^g(M_{\beta},L_{\nu})
\right.
\nn \\
&  & + 
\left.\frac{}{}
\left<N^{\rm c}_{\alpha}\right>_{P,s}
N^g(M_\alpha,L_{\mu})
\delta^{K}_{\alpha,\beta}\delta^{K}_{\mu,\nu}\right] \ \label{eq:PhiCov1} .
\ea
In order to proceed further we require an expression for the product
$\left< N^{\rm c}_{\alpha}N^{\rm c}_{\beta} \right>_{P,s}$. Again, this may be
obtained by following the arguments presented in
\citet{SmithMarian2011} (summarised in Appendix \ref{app:cluster}).
Thus we have,
\be \left< N^{\rm c}_{\alpha}N^{\rm c}_{\beta} \right>_{P,s}\equiv 
S_{\alpha\beta}+\Vs^2\overline{n}_{\alpha}\overline{n}_{\beta}+
  \Vs \overline{n}_{\alpha}\delta^{k}_{\alpha,\beta} \ . \label{eq:ClusCov}
\ee
The first term takes into account the excess variance above random in
the number counts, which arises due to the spatial correlations of the
clusters:
\be
S_{\alpha\beta} \equiv 
\Vs^2\,\overline{n}_{\alpha}\, \overline{n}_{\beta}\, 
\overline{b}_{\alpha}\,\overline{b}_{\beta}\, \sigma^2_V
\label{eq:SampCov} \ ,
\ee 
where in the above we have defined the effective bias of the clusters
in the $\alpha$th mass bin to be,
\be 
\overline{b}_{\alpha} = 
\frac{1}{\overline{n}_{\alpha}} 
\int_{M_{\alpha}-\Delta M_{\alpha}/2}^{M_{\alpha}+\Delta M_{\alpha}/2} dM b(M) n(M)\ 
\label{eq:def_bias1}
\ee
and also introduced the volume variance
\be \sigma^2_V \equiv \int \frac{\dk}{(2\pi)^3} \left|W(\bk)\right|^2
P(k) \ , \label{eq:VolVar}\ee
where $W(\bk)$ is the survey window function and $P(k)$ is the matter
power spectrum. 

Substituting \Eqn{eq:ClusCov} into \Eqn{eq:PhiCov1}, gives
\ba \left<\widehat{\phi}_\mu\widehat{\phi}_\nu\right> & = &
\frac{1}{\Delta L_{\mu}\Delta L_{\nu}\Vs^2} \left\{ \sum_{\alpha=1}^{N_M}
\sum_{\beta=1}^{N_M} N^g(M_\alpha,L_{\mu}) \right. \nn \\ 
& &  \times \ N^g(M_{\beta},L_{\nu}) 
\left[\frac{}{}S_{\alpha\beta}+\Vs^2\overline{n}_{\alpha}\overline{n}_{\beta}+
  \Vs\overline{n}_{\alpha}\delta^{k}_{\alpha,\beta}\right] \nn \\ 
& & \left. + \sum_{\alpha=1}^{N_M}
\Vs\overline{n}_{\alpha}N^g(M_\alpha,L_{\mu})
\delta^{K}_{\alpha,\beta}\delta^{K}_{\mu,\nu} \right\} \ .\ea
Using \Eqnss{eq:SampCov}{eq:VolVar} in the above expression, gives
\ba 
\left<\widehat{\phi}_\mu\widehat{\phi}_\nu\right> & = & 
\frac{1}{\Delta L_{\mu}\Delta L_{\nu}}
\sum_{\alpha,\beta}
N^g(M_\alpha,L_{\mu})
N^g(M_{\beta},L_{\nu})
\nn \\
& & \frac{}{}\times \overline{n}_{\alpha}\overline{n}_{\beta} 
\left[\overline{b}_{\alpha}\overline{b}_{\beta}\sigma^2_V+1\right]
 \nn \\
& & + \frac{1}{\Delta L_{\mu}\Delta L_{\nu}\Vs}\sum_{\alpha}
\overline{n}_{\alpha}
N^g(M_\alpha,L_{\mu})
N^g(M_{\alpha},L_{\nu}) \nn \\
& & +\frac{1}{\Delta L_{\mu}\Delta L_{\nu}\Vs}\sum_{\alpha=1}^{N_M}
\overline{n}_{\alpha} N^g(M_\alpha,L_{\mu})
\delta^{K}_{\mu,\nu} \ . \label{eq:inter}
\ea
Again, if the mass bins are sufficiently narrow, then we may use the
mean value theorem to make the following approximations:
$\overline{n}_{\alpha}\approx n(M_{\alpha})\Delta M_{\alpha}$ and
$\overline{b}_{\alpha}\approx b(M_{\alpha})$. This allows us to
transform the above expression into integrals over cluster mass. Next,
if we subtract off the second term on the right hand side of
\Eqn{eq:LFcov}, this gives us the covariance matrix of the GLF. Note,
that this simply removes the $+1$ from the first term in square
brackets in \Eqn{eq:inter}. Thus we find,
\ba 
{\mathcal C}_{\mu\nu} & = & 
\frac{1}{\Delta L_{\mu}\Delta L_{\nu}} \left\{
\int dM_1 \int dM_2 n(M_1)n(M_2) \right.  \nn \\
& & \times \ b(M_1)b(M_2)\sigma^2(\Vs)
N^g(M_1,L_{\mu})
N^g(M_2,L_{\nu})\nn \\
& & + \frac{1}{\Vs}
\int dM_1 n(M_1)
N^g(M_1,L_{\mu})
N^g(M_1,L_{\nu}) \nn \\
& & + \left. \frac{1}{\Vs}\int dM_1 n(M_1)
N^g(M_1,L_{\mu}) \delta^{K}_{\mu,\nu} \right\} \ . \label{eq:gram}
\ea
The above expression may be written in a more compact way by
introducing the following expressions: the effective bias of galaxies
in luminosity bin $L_\mu$,
\be
b^{\g}_\mu \equiv b^{\g}_{\mu}(L_{\mu}) \equiv  
\frac{1}{\bar{n}^{\g}_{\mu}}
\int dM_1 n(M_1) b(M_1)N^{\g}(M_1,L_{\mu})\ \ , \label{eq:lumbias}
\ee
and the effective number density of galaxies in the luminosity bin
$L_{\mu}$,
\be \bar{n}^g_{\mu} \equiv \bar{n}^g(L_{\mu}) \equiv 
\int dM_1 n(M_1) N^g(M_1,L_{\mu}) \ .
\ee
On using these definitions in \Eqn{eq:gram}, we find:
\ba 
& & \hspace{-0.5cm}
{\mathcal C}_{\mu\nu}=
\phi(L_{\mu})\phi(L_{\nu})
b^{g}(L_{\mu}) b^{g}(L_{\nu}) \sigma^2(\Vs)
+\frac{\phi(L_{\mu})  \delta^{K}_{\mu,\nu}}{\Vs \Delta L_{\mu}} 
\nn \\
& & + \frac{1}{\Vs}
\int dM_1 n(M_1)
\frac{N^g(M_1,L_{\mu})}{\Delta L_{\mu}}
\frac{N^g(M_1,L_{\nu})}{\Delta L_{\nu}}
\ea
Finally, we may reexpress our result in terms of the CLF of galaxies
$\Phi(L_{\mu}|M)$, as
\be 
{\mathcal C}_{\mu\nu} =  
\phi_{\mu}\phi_{\nu}
b^{g}_{\mu}\, b^{g}_{\nu}\, \sigma^2_V
+\frac{\phi_{\mu}  \delta^{K}_{\mu,\nu}}{\Vs \Delta L_{\mu}}  +  \Sigma_{\mu\nu}\ ,
\label{eq:CovVolLim}\ee
where we defined the \emph{`halo occupancy covariance'} to be
\be \Sigma_{\mu\nu}\equiv\frac{1}{\Vs}
\int dM_1 n(M_1) \Phi(L_{\mu}|M_1) \Phi(L_{\nu}|M_1)\ .\ee
Closer inspection of \Eqn{eq:CovVolLim} reveals several interesting
points. The first term informs us that the presence/absence of
large-scale structures in the survey volume will enhance/suppress the
number of galaxies in our estimates and that this will lead to
bin-to-bin correlations in the estimates of the GLF. The second term
is the standard Poisson error term, which dominates in the limits of
rare counts. The third term is interesting, and tells us that, if our
understanding of galaxy formation is correct and galaxies only appear
inside haloes, then, even in the absence of structure, GLF estimates
are correlated. This owes to the fact that, if we have a halo, then it
most likely comes with a set of $\phi(L|M)$ galaxies and so the
presence of one galaxy is correlated with the presence of additional
galaxies. Finally, we note that \citet{Robertson2010} wrote down terms
similar to the first two in our \Eqn{eq:CovVolLim}. However, owing to
his over-simplistic model for the number of galaxies hosted by a halo
of a given mass, he failed to obtain the halo occupancy covariance
term.


\subsection{Luminosity function correlation matrix}\label{ssec:corr}

A short corollary to this section is that we may now construct the
correlation matrix from the covariance matrix:
\be r_{\mu\nu}\equiv \frac{C_{\mu\nu}}{\sqrt{C_{\mu\mu}C_{\nu\nu}}}\ . \ee
This obeys the inequality $|r_{\mu\nu}|\le1$. 

Inserting our expression for the covariance matrix given by
\Eqn{eq:CovVolLim} into the above definition, we find
\ba 
r_{\mu\nu}  & = & 
\frac{\phi_{\mu}\phi_{\nu}b^{g}_{\mu} b^{g}_{\nu} \sigma^2(\Vs)
+\frac{\phi_{\mu}  \delta^{K}_{\mu,\nu}}{\Vs \Delta L_{\mu}} + \Sigma_{\mu\nu}}
{\prod_{i=\{\mu,\nu\}}\left[\phi_{i}^2[b^{g}_{i}]^2 \sigma^2(\Vs)
+\frac{\phi_{i}}{\Vs \Delta L_{i}} + \Sigma_{ii}\right]^{1/2}}\ \label{eq:cc1} \ .
\ea
Let us now factor out the Poisson error terms from the numerator and
denominator of \Eqn{eq:cc1}. Note that the term in the numerator may
be rewritten as
\be \frac{\phi_{\mu} \delta^{K}_{\mu,\nu}}{\Vs\Delta
L_{\mu}} = \frac{\sqrt{\phi_{\mu}\phi_{\nu}} }
{\sqrt{\Vs\Delta L_{\mu}\Vs \Delta L_{\nu}}}\delta^{K}_{\mu,\nu}\ .\ee 
Whereupon,
\ba 
r_{\mu\nu}  & = & 
\frac{\sqrt{N^{g}_{\mu}N^{g}_{\nu}}b^{g}_{\mu} b^{g}_{\nu} \sigma^2(\Vs)
+ \widetilde{\Sigma}_{\mu\nu} + \delta^{K}_{\mu,\nu}}
{\prod_{i=\{\mu,\nu\}}\left[N^{g}_{i}(b^{g}_{i})^2 \sigma^2(\Vs)
+ \widetilde{\Sigma}_{ii} +1
\right]^{1/2}} \label{eq:crosscorr}
\ea
and in the above we have defined the total number of surveyed galaxies
in the luminosity bin $L_{\mu}$ to be, \mbox{$N^{g}_{\mu}\equiv
  \phi_{\mu}\Delta L_{\mu}\Vs$}, and where we have defined:
\be 
\widetilde{\Sigma}_{\mu\nu} \equiv  
\frac{\sqrt{\Vs\Delta L_{\mu}\Vs\Delta L_{\nu}}}{\sqrt{\phi_{\mu}\phi_{\nu}}}\Sigma_{\mu\nu}\ .
\label{eq:sigtilde}
\ee
On manipulating the above expression, we find that it may also be
written as,
\be 
\widetilde{\Sigma}_{\mu\nu}=
\frac{\int dM n(M)N(L_{\mu}|M)N(L_{\nu}|M)}{\prod_{i=\{\mu,\nu\}} \left\{ \int dM n(M)N(L_i|M) \right\}^{1/2}}\ .
\ee

Several cases of interest may be noted. If, for the moment, we neglect
the halo occupancy covariance, i.e
$\widetilde{\Sigma}_{\mu\nu}\rightarrow0$, then we note the two cases:
\ba 
\sqrt{N^{g}_{\mu}N^{g}_{\nu}}b^{g}_{\mu} b^{g}_{\nu}
\sigma^2(\Vs)  & \ll & 1 \ ;\\
\sqrt{N^{g}_{\mu}N^{g}_{\nu}}b^{g}_{\mu} b^{g}_{\nu}
\sigma^2(\Vs) & \gg & 1 \ .
\ea
In the first, the errors are dominated by the Poisson sampling of the
galaxies and the covariance matrix is uncorrelated. In the second
case, the matrix is dominated by the sample covariance, and the matrix
can become perfectly correlated:
\be
r_{\mu\nu}   =  
\frac{\sqrt{N^{g}_{\mu}N^{g}_{\nu}}b^{g}_{\mu} b^{g}_{\nu} \sigma^2(\Vs)}
{\prod_{i=\{\mu,\nu\}}\left[N^{g}_{i}(b^{g}_{i})^2 \sigma^2(\Vs)\right]^{1/2}} 
\rightarrow 1 \ .
\ee
We may also make the important point that, taking
$\Vs\rightarrow\infty$ and hence $\sigma(\Vs)\rightarrow 0$,
\emph{does not} guarantee that the correlation between different
luminosity bins is negligible. As the above equations clearly show, it
is the quantity $\Vs\sigma^2(\Vs)$ that is required to vanish for
negligible correlation to occur. Indeed, for a power-law power
spectrum, we would have that $\Vs\sigma^2(\Vs)\propto R^3
R^{-(3+n)}\propto R^{-n}$, which can only be made to vanish for
$n>0$. For CDM we have a rolling spectral index, and $n>0$ for
\mbox{$k\lesssim 0.01\kMpc$}, which implies that \mbox{$\Vs\gtrsim 0.5
  \Gpccube$} for the covariance to diminish with increasing volume.

On the other hand, if we now neglect the sample variance term,
i.e. $\sigma^2(\Vs)\rightarrow0$, then we have the two cases:
\ba 
\widetilde{\Sigma}_{\mu\mu} & \ll & 1 \ ;\\
\widetilde{\Sigma}_{\mu\mu} & \gg & 1  \ .
\ea
Thus through computing the quantity: 
\be \widetilde{\Sigma}_{\mu\mu}= \frac{\int dM
  n(M)N^2(L_{\mu}|M)}{\int dM n(M)N(L_\mu|M)}\ , \label{eq:sigtilde}
\ee
one can determine the relative importance of the halo occupancy
covariance term with respect to the Poisson errors. Notice also that
this is independent of the survey volume, and thus in principle sets
the lower limit for the magnitude of the bin-to-bin correlations of
the luminosity function data. In \S\ref{sssec:diagerr} we shall
explicitly evaluate this expression for a particular CLF model.


\section{Flux limited surveys}\label{sec:LFFluxLim}


\subsection{Expectation of estimator}


We now turn to the more complicated case of estimating the GLF in flux
limited surveys. Consider an observer at position $\bx_o$, if they
survey all galaxies down to an apparent magnitude depth of $m_{\rm
  lim}$, then the GLF may be obtained through use of estimator E2
given in \Eqn{eq:E2}. In terms of the quantities used in
Sec.~\ref{sec:LFVolLim}, this estimator may be expressed as:
\be 
\widehat{\phi}(L_{\mu}|\bx_o) = 
\frac{1}{\Vmu\Delta L_{\mu}}
\sum_{\alpha=1}^{N_M}
\sum_{i=1}^{N^{\rm c}_{\alpha}} N^{\rm g}_{i,\alpha,\mu} \Theta(\bx^{\rm c}_i-\bx_o|L_{\mu}) \ ,
\label{eq:E2B}
\ee
where $N^{\rm c}_{\alpha}$ and $N^{\rm g}_{i,\alpha,\mu}$ are as defined in
\Eqn{eq:E1B}. There are two new components in the above equation.  The
first modification is that we require a survey selection function
$\Theta$, which has the form:
\be \Theta(\bx_i|L_{\mu}) = 
\left\{
\begin{array}{ll}
1 & \left[|\bx_i|\le\chi_{\rm max}(L_{\mu})\right]\\
0 & \left[|\bx_i|>\chi_{\rm max}(L_{\mu})\right]
\end{array} 
\right. \ ,
\ee
where $\chi^{\rm max}(L_{\mu})$ is the maximum distance that a source
of luminosity $L_{\mu}$, or identically absolute magnitude $M$ (see
\Eqn{eq:MagLum} for the conversion), can be seen, given the apparent
magnitude limit of the survey $m_{\rm lim}$:
\be \chi_{\rm max}(L_{\mu})=10^{\left[m_{\rm lim}-M(L_{\mu})-25\right]/5} 
\ [\Mpc]\ .\label{eq:chimax}\ee
The second modification is that the survey volume now becomes
$\Vs\rightarrow V^{\rm max}(L_{\mu})\equiv\Vmu$, which is the maximum
volume that a galaxy with a luminosity $L_{\mu}$ could have been found
in, given the flux limit of the survey $m_{\rm lim}$. For a survey of
solid angle $\Omega_{\rm s}$, this can be written
\be \Vmu = \int^{\chi_{\rm max}(L_{\mu})}_{0} 
\frac{dV(\chi)}{d\chi}{d\chi} \ ,\ee
where $dV(\chi)$ is the comoving volume element out to comoving
geodesic distance $\chi(a)$.  In what follows we shall assume a flat
space-time geometry and so take the survey volume at luminosity
$L_{\mu}$ to be,
\be \Vmu = \frac{\Omega_{\rm s}}{3} \chi_{\rm max}^3(L_{\mu}) \ .\ee

The expectation of the GLF estimator can be written
\ba
\left<\widehat{\phi}(L_{\mu})\right> & = &
\frac{1}{\Vmu\Delta L_{\mu}}
\sum_{\alpha=1}^{N_M}
\left<\sum_{i=1}^{N^{\rm c}_{\alpha}} N^{\rm g}_{i,\alpha,\mu} \Theta(\bx^{\rm c}_i|L_{\mu})\right>_{g,P,s}
\nn \\
& & \hspace{-1.0cm} =  \frac{1}{\Vmu\Delta L_{\mu}} \sum_{\alpha=1}^{N_M} 
N^g(M_{\alpha},L_{\mu}) 
\left< \sum_{i=1}^{N_{\alpha}^{\rm c}} \Theta(\bx^{\rm c}_i|L_{\mu})\right>_{P,s}\label{eq:this}  \ ,
\ea
where in the above, for convenience, we have taken $\bx_o$ as the
origin of the coordinate system. The last factor in the above equation
simply gives the number of clusters in mass bin $\alpha$ that host
galaxies of luminosity $L_{\mu}$, which would be detected in the
survey volume. We shall define this as,
\be 
N^{\rm c}_{\alpha}(L_\mu) \equiv \sum_{i=0}^{N^{\rm c}_{\alpha}}\Theta(\bx^{\rm c}_i|L_\mu) \label{eq:fluxcounts} \ .
\ee
On averaging the above expression over the sampling distributions, we
find
\be
\left< N_{\alpha}^{\rm c}(L_{\mu})\right>_{P,s}  
=  \overline{n}_{\alpha} \Vmu  \ .
\ee
Substituting this back into \Eqn{eq:this} we arrive at the result:
\be \left<\widehat{\phi}(L_{\mu})\right> = 
\frac{1}{\Vmu\Delta L_{\mu}} \sum_{\alpha=1}^{N_M} 
N^g(M_{\alpha},L_{\mu}) \, \overline{n}_{\alpha}\Vmu \ .
\ee
In the limit of small mass bins, then we may approximate
$\overline{n}_{\alpha}\approx \Delta M_{\alpha} n(M_\alpha)$, and hence rewrite
the above expression in integral form as,
\be \left<\widehat{\phi}(L_{\mu})\right> = 
\int dM n(M) \Phi(L_{\mu}|M)\ ,
\ee
where again we have used
$\Phi(L_{\mu}|M)=N^g(M_{\alpha},L_{\mu})/\Delta L_{\mu}$. This agrees
with the estimator for the volume limited survey, and hence when
dealing with a flux-limited survey E2 is also formally an unbiased
estimator.


\subsection{Estimator covariance}


The covariance matrix of GLF estimator E2 can be written,
\be {\mathcal C}^{\rm FL}_{\mu\nu} =
\left<\widehat{\phi}_\mu\widehat{\phi}_\nu\right> -
\left<\widehat{\phi}_\mu\right> \left<\widehat{\phi}_\nu\right>
\ . \label{eq:LFcov2} \ee
Similar to our analysis for \Eqn{eq:part0}, let us focus on the first
term on the right hand side, and on inserting \Eqn{eq:E2B}, we find
that this can be written,
\ba 
\left<\widehat{\phi}_\mu\widehat{\phi}_\nu\right> 
& = & \frac{1}{\Delta L_{\mu}\Delta L_{\nu}\Vmu\Vnu}\sum_{\alpha=1}^{N_M}
\sum_{\beta=1}^{N_M} \nn \\
& & \hspace{-1cm}\times \left<\sum_{i=0}^{N^{\rm c}_{\alpha}}\sum_{j=0}^{N^{\rm c}_{\beta}}
\left< N^g_{i,\alpha,\mu}N^g_{j,\beta,\nu}\right>_g\Theta(\bx^{\rm c}_i|L_\mu)\Theta(\bx^{\rm c}_j|L_\nu)
\right>_{P,s}\ . \label{eq:part00}
\ea
The inner average over the galaxy population is given by
\Eqn{eq:galcov}, and after inserting this in to \Eqn{eq:part00} we
find,
\ba 
\left<\widehat{\phi}_\mu\widehat{\phi}_\nu\right> 
& = & \frac{1}{\Delta L_{\mu}\Delta L_{\nu}\Vmu\Vnu} \left\{  
\sum_{\alpha,\beta}^{N_M} N^g(M_\alpha,L_{\mu}) \right. \nn \\ 
& \times & 
N^g(M_{\beta},L_{\nu})
\left<\sum_{i=0}^{N^{\rm c}_{\alpha}}\sum_{j=0}^{N^{\rm c}_{\beta}}\Theta(\bx^{\rm c}_i|L_\mu)\Theta(\bx^{\rm c}_j|L_\nu)
\right>_{P,s}\  \nn \\
& + & 
\left. N^g(M_\alpha,L_{\mu})
\left<\sum_{i=0}^{N^{\rm c}_{\alpha}}\Theta(\bx^{\rm c}_i|L_\mu)
\right>_{P,s}\delta^{K}_{\mu,\nu}\right\} \ , \label{eq:part11}
\ea
where in obtaining the last term in the above expression we used the
fact that $\Theta^2(\bx^{\rm c}_i|L_\mu)=\Theta(\bx^{\rm c}_i|L_\mu)$. Using
\Eqn{eq:fluxcounts} we may rewrite the correlation of `observable'
clusters, that host galaxies in the luminosity bins $L_{\alpha}$ and
$L_{\nu}$ as,
\be
\left<\sum_{i=0}^{N^{\rm c}_{\alpha}}\sum_{j=0}^{N^{\rm c}_{\beta}}\Theta(\bx^{\rm c}_i|L_\mu)\Theta(\bx^{\rm c}_j|L_\nu)
\right>_{P,s} \equiv \left<N^{\rm c}_{\alpha}(L_\mu)N^{\rm c}_{\beta}(L_\nu)\right>_{P,s} \label{eq:A} \ .
\ee
The above expression may be evaluated in exactly the same way as
\Eqn{eq:ClusCov}, however in this case we must take into account that
the survey volume varies with the luminosity bin.  This leads us to
write,
\ba \left<N^{\rm c}_{\alpha}(L_\mu)N^{\rm c}_{\beta}(L_\nu)\right>_{P,s} & = &
S_{\alpha\beta}(L_{\mu},L_{\nu}) +
\Vmu\overline{n}_{\alpha}\Vnu\overline{n}_{\beta} \nn \\
& & + {\rm min}[\Vmu,\Vnu]\overline{n}_{\alpha}\delta^{k}_{\alpha,\beta} \label{eq:B} \ ,
\ea
where in the above we have defined the quantity
\be
S_{\alpha\beta}[L_{\mu},L_{\nu}] \equiv 
\Vmu\Vnu\ \overline{n}_{\alpha}\, \overline{n}_{\beta}\, 
\overline{b}_{\alpha}\,\overline{b}_{\beta}\, \sigma^2(L_{\mu},L_{\nu})
\label{eq:C} \ ;
\ee 
with
\be 
\sigma^2(L_{\mu},L_{\nu}) \equiv \int \frac{\dk}{(2\pi)^3}
P(k)W(k|L_{\mu})W(k|L_{\nu}) \ .
\ee
The quantity $W(k|L_{\mu})$ represents the Fourier transform of the
window function associated with the survey volumes for galaxies
of luminosity $L_{\mu}$. Explicitly, this is written:
\be W(k|L_{\mu}) \equiv \frac{1}{\Vmu} \int \dx
\exp[i\bk\cdot\bx]\Theta(\bx|L_{\mu}) \ .\ee
On inserting \Eqnss{eq:A}{eq:C} into \Eqn{eq:part11}, we find
\ba \left<\widehat{\phi}_\mu\widehat{\phi}_\nu\right> & = & 
\frac{1}{\Delta L_{\mu}\Delta L_{\nu}}
\sum_{\alpha,\beta}^{N_M} 
N^g(M_\alpha,L_{\mu}) N^g(M_{\beta},L_{\nu}) 
\overline{n}_{\alpha}\overline{n}_{\beta}
\nn \\ 
& & \times
\left[\frac{}{}\overline{b}_{\alpha}\overline{b}_{\beta}\sigma^2(L_{\mu},L_{\nu})+1\right] \nn \\
& + & 
\frac{1}{\Delta L_{\mu}\Delta L_{\nu}}
\sum_{\alpha}^{N_M} 
\frac{{\rm min}[\Vmu,\Vnu]}{\Vmu\Vnu} \nn \\
& & \times N^g(M_\alpha,L_{\mu}) N^g(M_{\alpha},L_{\nu}) \overline{n}_{\alpha}
 \nn  \\ 
& + & 
\frac{1}{\Delta L_{\mu}\Delta L_{\nu}\Vmu}
\sum_{\alpha}^{N_M} 
N^g(M_\alpha,L_{\mu})\overline{n}_{\alpha}\delta^{K}_{\mu,\nu} \ .\ea
On inserting the above expression into \Eqn{eq:LFcov2} gives the
covariance matrix of GLF estimates. Note that, the subtraction of the
terms $\left<\phi_{\mu}\right>\left<\phi_{\nu}\right>$ simply
corresponds to removing the $+1$ from the above expression. In the
limit of narrow mass bins we may approximate
$\overline{n}_{\alpha}\approx n(M_{\alpha})\Delta M_{\alpha}$, and the
above sums may also be converted in to integrals. Finally, on using
the relation $\Phi(L|M)\equiv N^g(M,L)\Delta L$, we find the
covariance matrix for flux-limited GLF estimates to be:
\be
{\mathcal C}^{\rm FL}_{\mu\nu} = \phi_{\mu}\,\phi_{\nu}\,
b^{g}_{\mu}\, b^{g}_{\nu}\, \sigma^2(L_{\mu},L_{\nu}) 
+\frac{\phi_{\mu}  \delta^{K}_{\mu,\nu}}{\Vmu \Delta L_{\mu}} 
+ \Sigma_{\mu\nu}^{\rm FL}\ . \label{eq:CovFluxLim}
\ee
In the above, we have defined the halo occupancy covariance matrix for
flux limited surveys to be:
\ba
\Sigma_{\mu\nu}^{\rm FL}\equiv \frac{1}{{\rm max}[\Vmu,\Vnu]}
\int dM n(M)
\Phi(L_{\mu}|M)\Phi(L_{\nu}|M)\ .\nn \\
\ea
Note, that if we take $\Vmu=\Vs$, then we exactly recover our earlier
result of \Eqn{eq:CovVolLim} for the volume limited sample.


\subsection{Luminosity function correlation matrix}\label{ssec:corrFL}

Following the discussion of \S\ref{ssec:corr} and from
\Eqn{eq:crosscorr}, we may write the correlation matrix for flux
limited surveys as:
\ba 
r^{\rm FL}_{\mu\nu}\!\!  & = & \!\!
\frac{\sqrt{N^{g}_{\mu}\,N^{g}_{\nu}}\,b^{g}_{\mu}\, b^{g}_{\nu} \sigma^2(L_{\mu},L_{\nu})
+ \widetilde{\Sigma}^{\rm FL}_{\mu\nu} + \delta^{K}_{\mu,\nu}}
{\prod_{i=\{\mu,\nu\}}\left[N^{g}_{i}(b^{g}_{i})^2 \sigma^2(L_{\mu},L_{\mu})
+ \widetilde{\Sigma}^{\rm FL}_{ii} +1
\right]^{1/2}}\ , \label{eq:crosscorrFluxLim}
\ea
where in the above we have defined the total number of surveyed
galaxies in the luminosity bin $L_{\mu}$ to be,
\mbox{$N^{g}_{\mu}\equiv \phi_{\mu}\Delta L_{\mu}\Vmu$}, and where
\be 
\widetilde{\Sigma}^{\rm FL}_{\mu\nu} \equiv \frac{\sqrt{\Vmu\Delta L_{\mu}\Vnu\Delta L_{\nu}}}
{\sqrt{\phi_{\mu}\phi_{\nu}}} \Sigma^{\rm FL}_{\mu\nu}\ .
\ee
Note that for the diagonal elements of this matrix, it can be shown
that $\widetilde{\Sigma}^{\rm
  FL}_{\mu\mu}=\widetilde{\Sigma}_{\mu\mu}$.  As before, several cases
of interest may be noted. Firstly, if we neglect the occupancy
variance, $\widetilde{\Sigma}^{\rm FL}_{\mu\nu}\rightarrow 0$, then we
have:
\ba 
\sqrt{N^{g}_{\mu}N^{g}_{\nu}}b^{g}_{\mu} b^{g}_{\nu}
\sigma^2(L_{\mu},L_{\nu}) & \ll & 1 \ ;\\
\sqrt{N^{g}_{\mu}N^{g}_{\nu}}b^{g}_{\mu} b^{g}_{\nu}
\sigma^2(L_{\mu},L_{\nu})  & \gg & 1  \ .
\ea
As in \S\ref{ssec:corr}, the first condition leads to Poisson
dominated counts and an uncorrelated matrix, and the second to a
perfectly correlated matrix. We thus deduce that for the case of the
flux limited survey, the sample covariance will only vanish when
$\sqrt{\Vmu\Vnu}\sigma^2(L_{\mu},L_{\nu})\rightarrow 0$.

Alternatively, for the case of no sample variance,
$\sigma^2(L_{\mu},L_{\nu})\rightarrow0$, we have the same situation as
for the volume limited case, and \Eqn{eq:sigtilde} provides an
indication of the relative strength of the halo occupation variance
with respect to the Poisson noise, which is independent of the survey
volume.

\vspace{0.2cm} In the following sections we will attempt to quantify
the level of covariance in volume limited and flux limited GLF
estimates.


\begin{figure*}
\centerline{
\includegraphics[width=8.5cm]{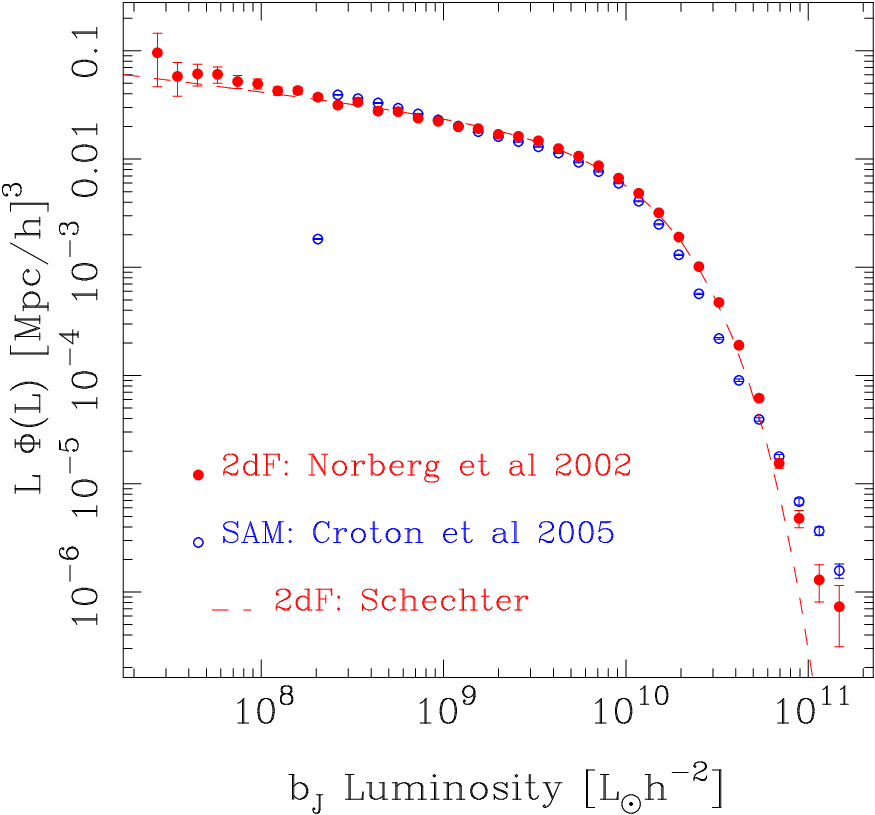}\hspace{0.3cm}
\includegraphics[width=8.5cm]{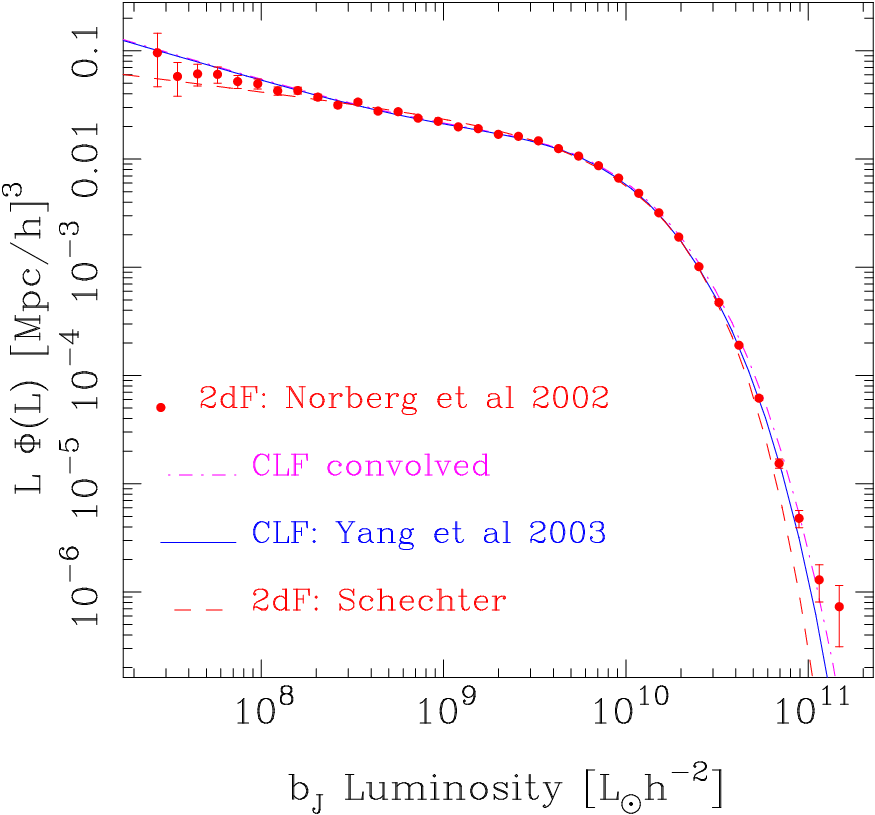}}
\caption{\small{The galaxy luminosity function in the 2dFGRS survey.
    In both panels, the solid red points with errors show the
    measurements from the 2dFGRS and the dashed red line denotes
    Schechter function fit from \citet{Norbergetal2002bshort}. {\em
      Left panel:} comparison of the 2dFGRS results with the
    luminosity function estimates made in \S\ref{ssec:samLF} from the
    semi-analytic model galaxy catalogue of \citet{Crotonetal2006}.
    {\em Right panel:} Comparison of the 2dFGRS results with the
    conditional luminosity function (CLF) model of
    \citet{Yangetal2003}, denoted by the solid blue line. Note that
    the magenta dot dashed line shows the effect of convolving the CLF
    model magnitudes with the lognormal magnitude error model for the
    2dFGRS as described by \citet{Norbergetal2002bshort}. }
\label{fig:LF}}
\end{figure*}


\section{Empirical results and modelling}\label{sec:models}

In this section we briefly summarise the procedures that we use for
modelling the GLF, the CLF and the luminosity dependence of galaxy
clustering.


\subsection{An empirical luminosity function}

Over the past few decades, the Schechter function has been found to
provide a reasonably good description of the GLF.
\be \phi(L) \dL =\phi_* \left(\frac{L}{L_*}\right)^{\alpha}\exp
\left[-\frac{L}{L_*}\right] \frac{\dL}{L_*}\ \ ,\label{eq:Schechter}\ee
where $L_{*}$ and $\phi_*$ are the characteristic luminosity and
number density of the surveyed galaxies, and $\alpha$ describes the
power-law slope of the faint galaxies. 

For the 2dFGRS survey, the best fit Schechter function parameters, for
galaxies with $K$-corrected $b_j$ luminosities in the range $(-22.5\le
M_{b_j}-5\log_{10}h\le-14.0)$, were found to be
\citep{Norbergetal2002bshort}: \mbox{$L_{*}=9.64\times
  10^9\Lsol $}, $\alpha=-1.21$ and
$\phi_*=1.61\times10^{-2}h^3{\rm Mpc}^{-3}$.


\subsection{Luminosity function of semi-analytic galaxies}\label{ssec:samLF}

As was shown by \citet{Kauffmannetal1999} and \citet{Coleetal2000},
semi-analytic models (SAM) of galaxy formation are a promising way to
attempt to understand the complex physics of galaxy formation. The
main advantage of this approach is that it allows one to rapidly
explore the effects of physical scaling relations on the observational
properties of galaxies. This property also makes it a useful tool for
generating mock galaxy catalogues.

In this study we make use of the publicly available SAM catalogues of
\citet{Crotonetal2006}. These model galaxies were generated by
carefully following the detailed merger histories of dark matter
haloes within the Millennium Run $N$-body simulation. This was an
$N$-body simulation that followed the non-linear evolution of
structure formation with $N=2048^3$ dark matter particles in a cubical
box of length $L=500\Mpc$. The cosmological model for this simulation
was: $\{\Omega_{\rm
  m}=0.25,\,\Omega_\Lambda=0.75,\,n_s=1.0,\,\sigma_8=0.9,\,h=0.73\}$,
where these are the matter and vacuum energy density parameters, the
primordial power-spectral index, the power spectrum normalisation, and
the dimensionless Hubble parameter, respectively \citep[for full
details see][]{Springeletal2005}. Through a novel treatment of AGN
feedback in the radio spectrum, the authors were able to show that the
predicted bright end of the GLF could be qualitatively reconciled with
observations from the 2dFGRS.

In these catalogues, galaxy magnitudes are available in both $BVRIK$
(Vega) or $ugriz$ (AB SDSS) filters.  Owing to the limited resolution
of the Millennium Run simulation, the SAM galaxies were only able to
be correctly followed down to $M_{b_J}-5\log_{10}h<-15.6$,
($L_{B}\sim2\times10^{8}\Lsol$). The 2dFGRS sample goes one order
of magnitude fainter. Having said that, the catalogues include a total
of about 9 million galaxies in the full simulation box, roughly
$\sim40$ times more mock galaxies than can be found in the 2dFGRS.

Figure~\ref{fig:LF}, left panel, compares the 2dFGRS $b_{J}$ GLF with
the GLF estimates obtained using \Eqn{eq:E1} for the SAM
galaxies. From this it can be seen that the \citet{Crotonetal2006}
galaxies do indeed provide a qualitatively good description of the
2dFGRS GLF. The largest deviations are noticeable for galaxies with
$L>L_{\star}$. Also, we see the drop-off in the number density of
objects present around $L\sim2\times10^8\Lsol$, due to the limited
mass resolution of the simulation.


\subsection{Conditional luminosity function}\label{ssec:clf}

The CLF was first introduced by \citet{Yangetal2003}. We now summarise
their model, which has been highly successful at reproducing a large
number of observational results from the 2dFGRS. Again, given a halo
of mass $M$, the CLF returns the number of galaxies per unit
luminosity interval $\dL$. It can be represented by a Schechter type
function:
\be \Phi(L|M) \dL =\tilde{\Phi}_*
\left(\frac{L}{\tilde{L}_*}\right)^{\tilde{\alpha}} \exp
\left[-\frac{L}{\tilde{L}_*}\right] \frac{\dL}{\tilde{L}_*}  \ ,
\ee
where the three free parameters
$\tilde{\Phi}_*\equiv\tilde{\Phi}_*(M)$,
$\tilde{L}_*\equiv\tilde{L}_*(M)$ and
$\tilde{\alpha}\equiv\tilde{\alpha}(M)$ are all mass dependent
quantities.  These parameters in turn are described by the following
functions:
\ba
\tilde{\alpha} & = & \frac{}{}\alpha_{15}+\eta\log_{10}(M/[10^{15}\Msol])\ ;  \\
\tilde{L}_* & = & 
\frac{2M}{f(\tilde{\alpha})}
\left[
\left(\frac{M}{L}\right)_0
\right]^{-1}
\left[
\left(\frac{M}{M_1}\right)^{-\beta}+
\left(\frac{M}{M_2}\right)^{\gamma_2}
\right]^{-1} \ ; \\
\tilde{\Phi}_{*} & = & 
\frac{\left<L\right>(M)}{\tilde{L}_{*}\Gamma[\tilde{\alpha}+1]}\ .
\ea
With the additional auxiliary functions:
\be \left<L\right>(M)=2M \left[
\left(\frac{M}{L}\right)_0
\right]^{-1}
\left[
\left(\frac{M}{M_1}\right)^{-\beta}+
\left(\frac{M}{M_1}\right)^{\gamma_1}
\right]^{-1} \ ;
\ee
\be
f(\tilde{\alpha})=\frac{\Gamma[\tilde{\alpha}+2]}{\Gamma[\tilde{\alpha}+1,1]}\ ,
\ee
where $\Gamma[x]$ and $\Gamma[x,a]$ are the Gamma and incomplete Gamma
functions, respectively: 
\ba
\Gamma[x] & = & \int_{0}^{\infty} dz z^{x-1}\exp(-z)\ ; \\
\Gamma[x,a] & = & \int_{a}^{\infty} dz z^{x-1}\exp(-z)\ .
\ea

There are 8 free parameters in the above model, these are augmented by
one final parameter $M_{\rm min}$, which specifies the minimum mass
halo that may host a galaxy. For these 9 parameters we use the
best-fit values reported in \citet{Yangetal2003}: ${\bf
  p}=\{\alpha_{15}=-1.32,\ \eta=-0.36,\ \log_{10} M_1=10.42,\
\log_{10} M_2=11.74,\ (M/L)_0=102,\ \beta=0.6,\ \gamma_1=0.28, \
\gamma_2=0.69,\ \eta=-0.36,\ \log_{10}M_{\rm min}=9.0\}$. 

Finally, the GLF can be obtained from the CLF, by integrating over the
halo mass function as described in \Eqn{eq:LFfromCLF}. Note, in
evaluating \Eqn{eq:LFfromCLF}, we follow exactly the recipe presented
in \citet{Yangetal2003} and adopt the \citet{ShethTormen1999} mass
function and the transfer function of
\citet{Bardeenetal1986}\footnote{Note, in evaluating the CLF model, we
  adopt the cosmological parameters used by \citet{Yangetal2003}:
  $\{\Omega_{\rm
    m}=0.3,\,\Omega_\Lambda=0.7,\,n_s=1.0,\,\sigma_8=0.9\}$. These are
  slightly different from those used in the Millennium simulation,
  however they are the same as those used in the estimation of the
  2dFGRS GLF.}.  This is necessary, since if we were to adopt other
models, then we would expect the quoted parameters to no longer be the
maximum likelihood parameter set. Since this is a first calculation we
are not too worried by this, however for a more precise calculation
one should reoptimise ${\bf p}$ for the true cosmological model.

Figure~\ref{fig:LF}, right panel, compares the 2dFGRS $b_{J}$ GLF with
the GLF obtained from \Eqn{eq:LFfromCLF}. As can be seen from the
figure, the CLF model of \citet{Yangetal2003} (solid blue line)
qualitatively provides a good description of the 2dFGRS data. Note,
the optimised best-fit parameters described in the paper of Yang et
al., do not take into account the presence of magnitude errors in the
2dFGRS data.  If we convolve the model magnitudes with the log-normal
distribution described in \citet{Norbergetal2002bshort}, then we find
a small increase in the abundance of the brightest galaxies.
Appendix~\ref{sec:magerrs} describes the inclusion of magnitude
errors.


\subsection{Luminosity dependence of galaxy clustering}

In order to make predictions for the covariance matrix of the GLF we
must also understand the luminosity dependence of the bias of the
galaxy distribution. We explore this using the SAM galaxies
\citep{Crotonetal2006}. First, the galaxy catalogue is sliced into 8
bins in absolute magnitude. The exact magnitude bins that we employ
and the numbers of galaxies in each bin are presented in
Table~\ref{tab:lumbins}. The correlation functions for the SAM
galaxies were then estimated using our parallel tree-code correlation
function algorithm {\tt DualTreeTwoPoint}, which is based on the
kD-Tree approach of \citet{MooreAetal2001}. The correlation functions
were estimated in 40 logarithmically spaced bins in the radial
interval $r\in[0.05,50.0]\Mpc$.


\begin{figure}
\centerline{
\includegraphics[width=8.5cm]{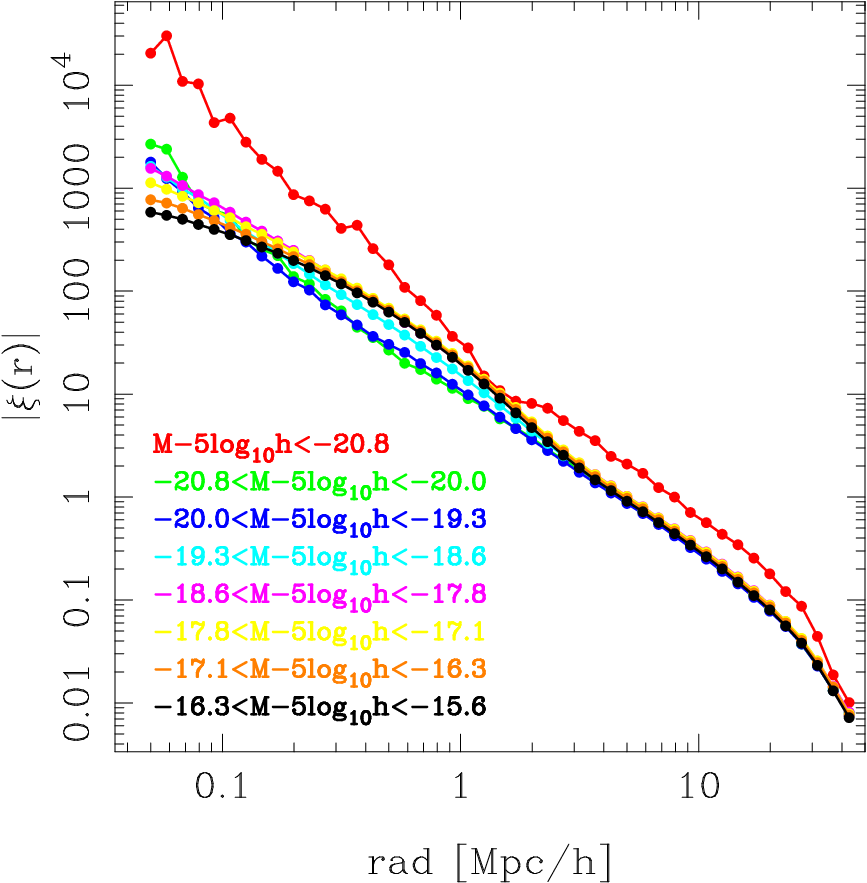}}
\caption{\small{Correlation functions estimated from the
    \citet{Crotonetal2006} semi-analytic galaxy catalogue in the
    Millennium simulation as a function of radial separation. The
    different coloured solid symbols show the results for the 8
    absolute magnitude bins described in Table~\ref{tab:lumbins}}.
\label{fig:LumCorr}}


\vspace{0.3cm}

\centerline{
\includegraphics[width=8.5cm]{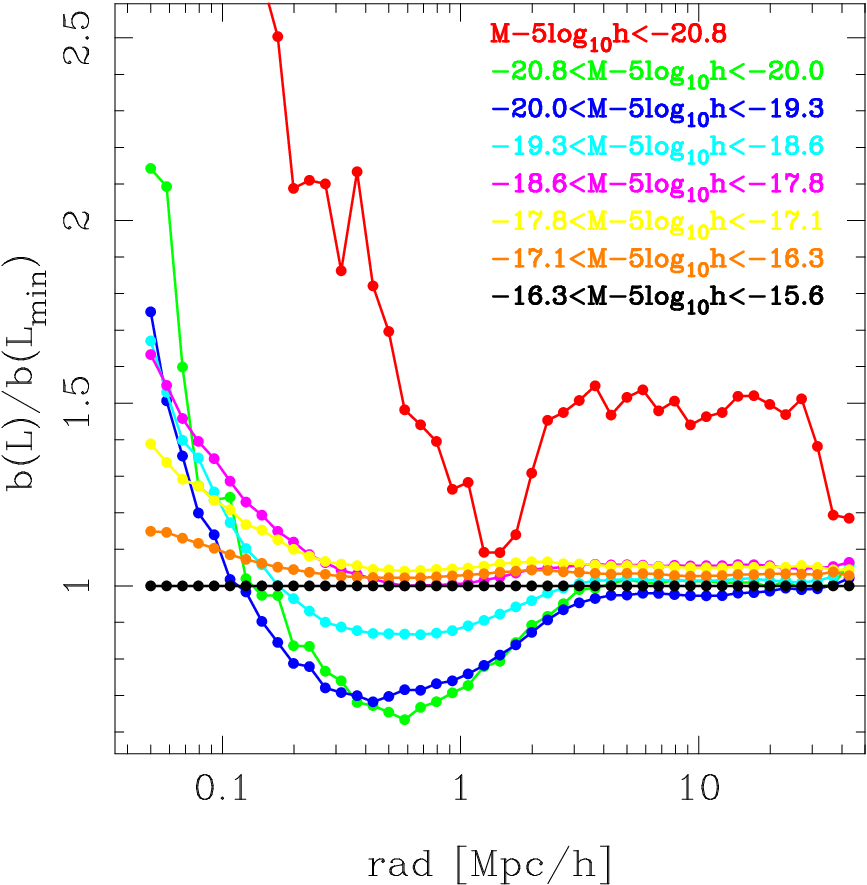}}
\caption{\small{Relative scale dependence of galaxy bias measured for
    the different galaxy populations in the Millennium simulation
    semi-analytic galaxy catalogues of \citet{Crotonetal2006}. The
    relative bias is defined with respect to the lowest luminosity
    galaxy bin. The connected coloured points show the results for the
    8 magnitude bins presented in Table~\ref{tab:lumbins}.}
\label{fig:LumBias}}
\end{figure}


\begin{table}
\centering{
\caption{Table showing: Col.~1: bin number; Col~2: the absolute
  magnitude limits of the bin; Col.~3: Number of galaxies within the
  bin from which we calculate the correlation functions.}
\label{tab:lumbins}
\begin{tabular}{c|rr}
\hline 
Bin    & Magnitude range        & Number of \\ 
Number & $[M_{b_{\rm J}}-5\log_{10}h]$  & Galaxies  \\ 
\hline 
1  & $[<-20.8]$      & 15,448 \\
2  & $[-20.0,-20.8]$ & 130,447\\
3  & $[-19.3,-20.0]$ & 471,467\\
4  & $[-18.6,-19.3]$ & 876,150\\
5  & $[-17.8,-18.6]$ & 125,4400 \\
6  & $[-17.1,-17.8]$ & 1,690,406 \\
7  & $[-16.4,-17.1]$ & 2,367,636 \\
8  & $[-15.6,-16.4]$ & 3,119,262 
\end{tabular}}
\end{table}


Figure~\ref{fig:LumCorr} shows the results for the galaxy correlation
functions measured in the 8 luminosity bins presented in
Table~\ref{tab:lumbins}. On scales $r>3\Mpc$ the signal appears to
demonstrate a power-law like form and with the brightest sample of
galaxies being significantly more correlated than the lower luminosity
galaxies. On smaller scales, however the signal is more complex: there
appears to be a strong scale-dependence with lower luminosity galaxies
becoming more strongly correlated than intermediate luminosity
galaxies.

Figure~\ref{fig:LumBias} quantifies this scale-dependence in more
detail, where we plot the relative bias of the SAM galaxies as a
function of scale. We define the relative galaxy bias as:
\be 
b_{\rm rel}(L_{\mu},L_{\nu})\equiv \frac{b^{\g}(L_{\mu})}{b^{\g}(L_{\nu})} = 
\sqrt{\frac{\xi_{\rm gg}(r|L_\mu)}{\xi_{\rm gg}(r|L_{\nu})}} \ .
\ee
The figure shows $b^{\g}_{\rm rel}(L_{\mu},L_{\rm min})$. On scales
$r>3\Mpc$ the bias is reasonably flat for all of the bins, but that,
interestingly, the lower luminosity galaxies can be more strongly
correlated than the intermediate luminosity bins.  Furthermore, it
shows that on scales less than $r\sim1\Mpc$ the brightest galaxy bins,
with the exception of Bin 1, all possess a strong relative anti-bias,
although Bin 1 does demonstrate a sharp dip at about the same
scale. On still smaller scales, $r<100\kpc$, the relative bias becomes
strongly positive.  Owing to the fact that we are primarily interested
in understanding the luminosity dependence of the large-scale bias, we
shall reserve the understanding of this scale-dependence for future
work.

We now focus on the large-scale relative bias. In most observational
studies the relative bias is computed with respect to the
characteristic luminosity $L_{*}$ of the survey. For the SAM galaxies,
this approximately corresponds to the galaxies in Bin 3. For this
work, our operational definition of `large scales' is given by
$\left(5\Mpc<r<30 \Mpc\right)$.

\begin{figure}
\centerline{
\includegraphics[width=8.5cm]{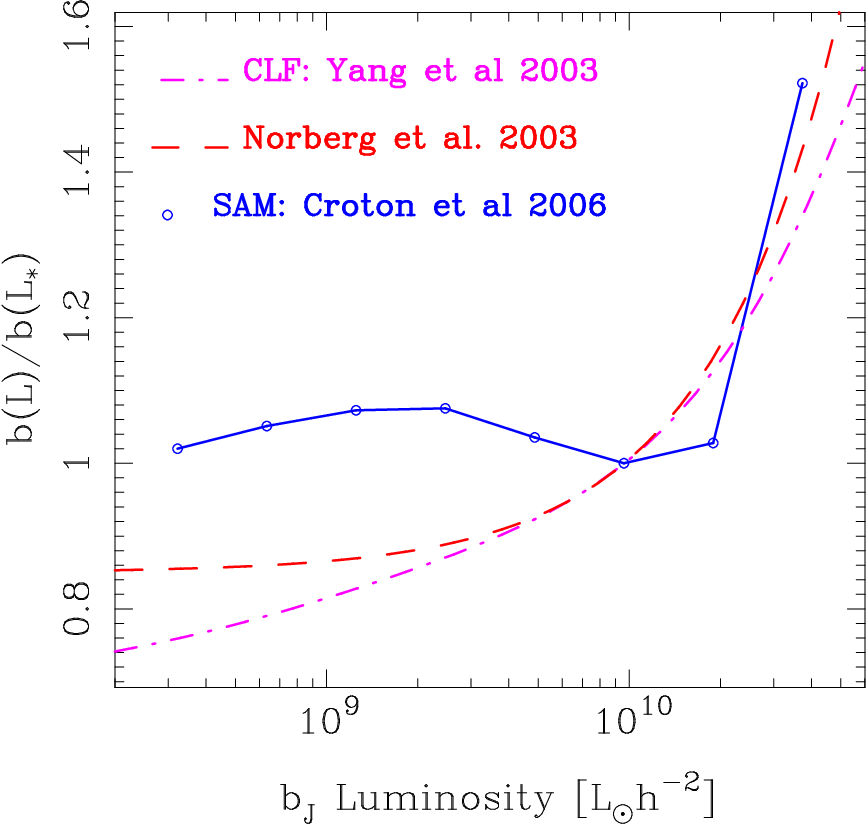}}
\caption{\small{Relative large-scale bias of galaxies as a function of
    luminosity. The dashed red line represents the results from the
    2dFGRS presented in \citet{Norbergetal2002ashort}; the connected
    open blue points denote our estimates from the semi-analytic
    galaxy catalogues in the Millennium simulation
    \citep{Crotonetal2006}; and the magenta dot-dashed line denotes
    the results from the conditional luminosity function model of
    \citet{Yangetal2003}, which were constrained to match the 2dFGRS
    results.}
\label{fig:LFbiasComp}}
\end{figure}

Figure~\ref{fig:LFbiasComp} shows $b^{\g}_{\rm
  rel}(L_{\mu},L_{\nu}={\rm Bin\, 3})$, and the luminosity dependence
of the large-scale relative bias measured from the SAM galaxies is
represented by the blue solid line. This may be compared with the
results for the 2dFGRS obtained by \citet{Norbergetal2002ashort}:
\be 
\frac{b_{\rm 2dF}(L)}{b(L_{*})}= 0.85+0.15\left(\frac{L}{L_{*}} \right) \ ,
\ee
and represented in the figure by the red-dashed line.

Interestingly, we see that the relative bias for the SAM galaxies is
much flatter for faint objects than one finds for the 2dFGRS. The
relative bias appears to have a minimum for $L_{*}$ galaxies and then
increases slightly for fainter objects, whereas the bias steadily
decreases for the 2dFGRS.  However, the SAM galaxies do correctly
capture the trend that the brightest galaxies in the 2dFGRS are more
strongly correlated than the fainter ones.  Thus, whilst the SAM
galaxies are able to reproduce the GLF, they appear to only
qualitatively capture the luminosity dependence of the clustering in
the 2dFGRS. This failure of the \citet{Crotonetal2006} model to
correctly capture the luminosity dependence of the clustering has
been noted in previous studies
\citep{Lietal2007,Kimetal2009,Guoetal2011}. These have attributed the
discrepancy between the observations and the model to the fact that,
too many faint satellite galaxies are placed in the high mass haloes.

We may also obtain a prediction for the luminosity dependence of the
bias using the CLF approach of \citet{Yangetal2003}.  On rewriting
\Eqn{eq:lumbias} in terms of the CLF we find,
\be
b^{g}(L_{\mu}) =
\frac{1}{\phi(L_\mu)}
\int dM_1 n(M_1) b(M_1)\Phi(L_{\mu}|M)\ \label{eq:lumdepbiasCLF} .
\ee
We have evaluated the above integral using the model described in
\S\ref{ssec:clf}, and the results are represented by the dot-dashed
line in \Fig{fig:LFbiasComp}. Clearly, this model appears to
accurately reproduce the luminosity dependence of the
clustering. However, this fact is not too remarkable, since the model
was optimised using this data. The salient point is that we are able
to reproduce the 2dFGRS results through evaluating
\Eqn{eq:lumdepbiasCLF}.


\section{Covariance of the galaxy luminosity function}\label{sec:results}

In this section we test our theoretical model for the covariance
matrix of the GLF estimates. We start with the volume limited sample,
and then move on to the more complex scenario of the flux limited
sample.


\subsection{Results: Volume limited samples}\label{sec:results-sub1}

We use the SAM galaxy catalogues to construct an estimate of the
covariance matrix of the GLF in volume limited samples. We do this by
following the approach for computing the cluster count covariance,
which was described in \citet{SmithMarian2011}. Briefly, we take the
full volume of the Millennium simulation mock and slice it up into
$n^3$ cubical cells.  On taking $n=4$ we have 64 quasi-independent
sub-volumes of size $L=125\Mpc$. For each of these sub-volumes we
estimate the GLF in 27 equal logarithmically spaced luminosity bins
using \Eqn{eq:E1}. From these 64 we construct the covariance matrix
using the simple unbiased estimator:
\be 
\widehat{C}_{\mu\nu} = \frac{1}{n^3-1}\sum_{i=1}^{n^3}
\left[\widehat{\phi}_i(L_{\mu})-\overline{\phi}(L_{\mu})\right]
\left[\widehat{\phi}_i(L_{\nu})-\overline{\phi}(L_{\nu})\right] \label{eq:EstCorrVolLim}
\ee
where 
\be \overline{\phi}(L_{\mu}) =
\frac{1}{n^3}\sum_{i=1}^{n^3}\widehat{\phi}_i(L_{\mu}) \ .\ee

We are interested in exploring errors for a survey with volume
$V\sim0.125\Gpccube$, however, the above procedure provides us with
the covariance matrix for survey volumes of the order
$V=1.99\time10^{-3}\Gpccube$.  We obviate this problem by
approximating the covariance of the large volume to be the covariance
on the mean, i.e.
\be 
\widehat{C}_{\mu\nu}(V)\approx \widehat{\overline{C}}_{\mu\nu}(V/n^3) \equiv 
\widehat{C}_{\mu\nu}(V/n^3)/n^3 \ .
\ee

Furthermore, in order to make predictions from the theory we must
compute $\sigma^2_{V}$, i.e. \Eqn{eq:VolVar}. This requires us to
specify the survey window function. As described in
\citet{SmithMarian2011}, one must actually be quite careful when
computing this: if one wants to compare predictions with results from
simulations then one needs to use the exact density modes that are in
the box; if one wants to make predictions for the real Universe then
the simulations fail to capture this correctly when the box-length $L$
is comparable with the dimensions of the survey. In this case one
should use theoretical predictions. Since here we are comparing with
$N$-body simulations, a good approximation is to interpret the survey
volume as being spherical in the following way:
\be 
R=\left(\frac{3\Vs}{4\pi}\right)^{1/3}
\ee
and take the window function to be
\be W(k|R) = \frac{3}{y^3}\left[\sin y-y\cos y\right] \ ; \ y\equiv kR\ .\ee
Hence, the volume variance takes the simple form
\be \sigma^2_V = \int_{0}^{\infty} \frac{dk k^2}{2\pi^2} \left|W(k|R)\right|^2
P(k) \ . \label{eq:VolVar222222}\ee
%


\begin{figure}
\centerline{
\includegraphics[width=8.5cm]{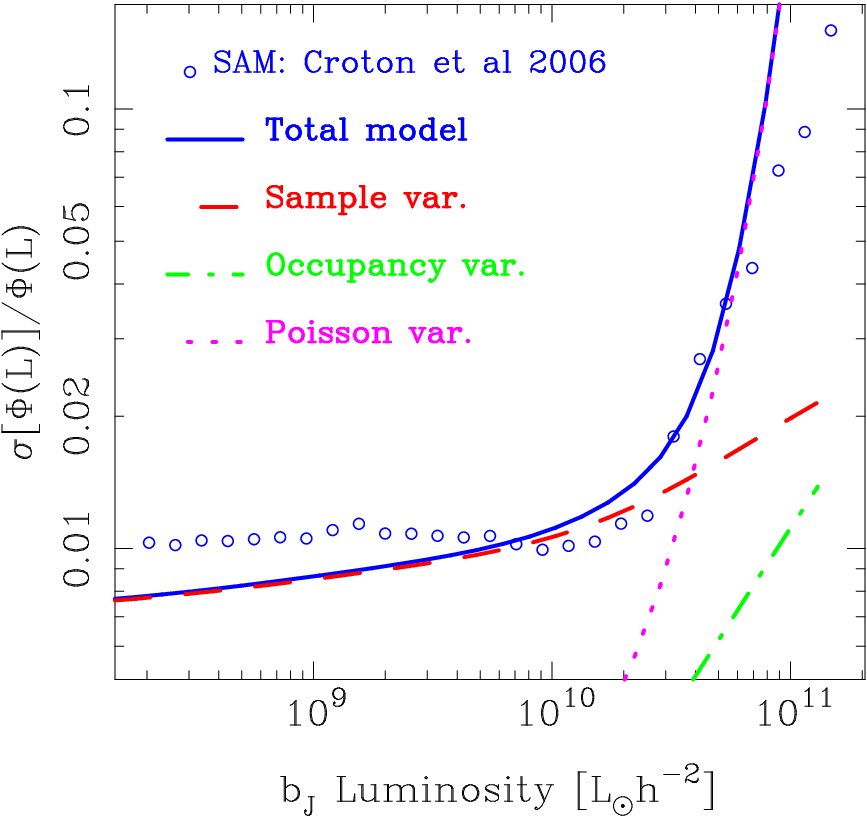}}
\caption{\small{Fractional errors in the galaxy luminosity function
    for a volume limited survey as a function of galaxy
    luminosity. The circular open points represent estimates obtained
    from the SAM galaxies. The solid blue line presents the total
    prediction of the theoretical model given by
    \Eqn{eq:CovVolLim}. The red dashed line denotes the contribution
    to the error from the sample variance; the green dot-dashed line
    corresponds to the error coming from the halo occupancy
    covariance; the magenta dotted line corresponds to the error from
    Poisson noise.}
\label{fig:LFDiagErr}}
\end{figure}


\subsubsection{Diagonal errors}\label{sssec:diagerr}

Figure~\ref{fig:LFDiagErr} shows the diagonal elements of the
covariance matrix divided by the ensemble average GLF estimates from
the 64 sub-cubes in the Millennium simulation (open points) as a
function of luminosity. In this figure, we also compare these results
with the theoretical predictions from \Eqn{eq:CovVolLim}, where we
have used the CLF model of \citet{Yangetal2003} as the model
input. From \Eqn{eq:CovVolLim} we find that,
\be 
\frac{\sigma^2[\phi(L_{\mu})]}{\phi^2(L_{\mu})}
= [b^{\rm g}_{\mu}]^2\sigma^{2}_{V}+
\frac{1}{N^{\rm g}_{\mu}}+\frac{\Sigma_{\mu\mu}}{\phi_{\mu}^2} \ .
\ee
The above expression informs us that: in the limit where the sample
variance is dominant, which for this case occurs when $N^{\rm g}_{\mu}>
10^4$, we have:
\be 
\left.\frac{\sigma[\phi(L_{\mu})]}{\phi(L_{\mu})}\right|_{\rm S.V.}
= b^{\rm g}_{\mu}\sigma_{V} \ ; \label{eq:SV}
\ee
and in the limit where the Poisson noise is dominant, which for this
case occurs when $N^{\rm g}_{\mu}< 10^4$, we have:
\be 
\left. \frac{\sigma[\phi(L_{\mu})]}{\phi(L_{\mu})}\right|_{\rm P.V.}
=  \frac{1}{\sqrt{N^{\rm g}_{\mu}}} \ .\label{eq:PV}
\ee
The luminosity dependence of the halo occupancy variance scales as
\be
\left. \frac{\sigma[\phi(L_{\mu})]}{\phi(L_{\mu})}\right|_{\rm O.V.} = 
\frac{\sqrt{\widetilde{\Sigma}_{\mu\mu}}}{\sqrt{N^{\rm g}_{\mu}}}
\label{eq:OV}
\ee
where $\widetilde{\Sigma}_{\mu\mu}$ is given by \Eqn{eq:sigtilde}. We
see from the denominator that this term scales in a similar way to the
Poisson noise. 



\begin{figure}
\centerline{
\includegraphics[width=8.5cm]{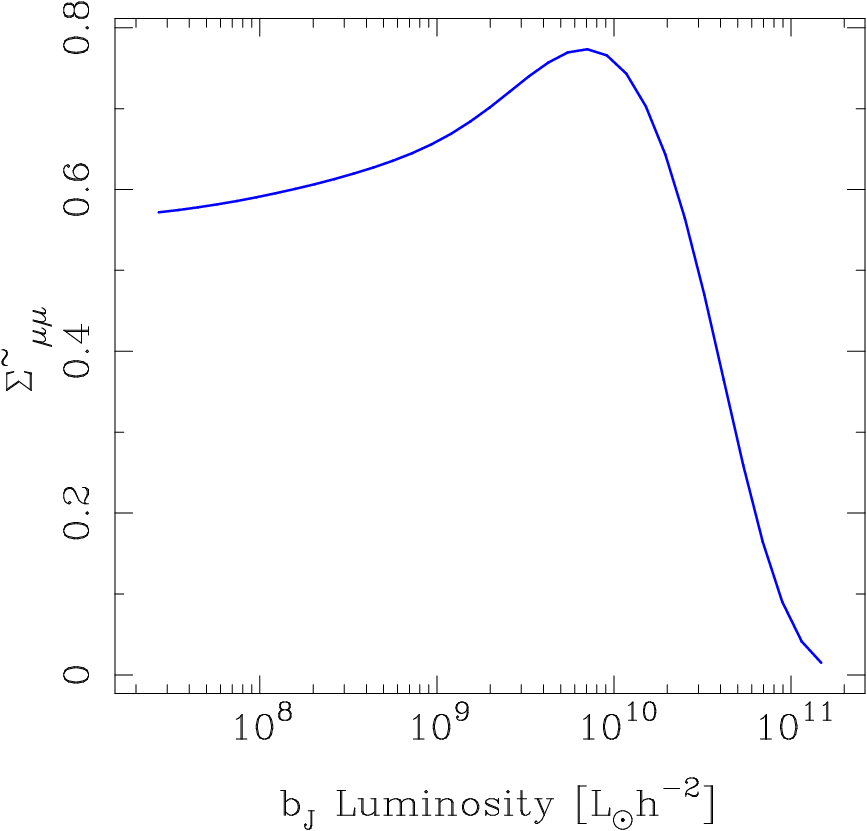}}
\caption{\small{Relative importance of the halo occupancy variance
    $\widetilde{\Sigma}_{\mu\mu}$ (c.f.~\Eqn{eq:sigtilde}) with
    respect to the Poisson errors, as a function of the luminosity
    bin. Poisson errors dominate as
    $\widetilde{\Sigma}_{\mu\mu}\ll1$. Note that this is independent
    of survey volume.}
  \label{fig:relerr}}
\end{figure}


Figure~\ref{fig:LFDiagErr} shows that the theory and the measurements
from the SAM are in excellent agreement. Further the above limiting
cases are clearly demonstrated by the data. Note that for
low-luminosities the errors in the SAM data appear to be slightly in
excess of the theoretical predictions. From \Eqn{eq:SV} we see that
this can be attributed to the fact that the luminosity dependence of
the bias in the SAMs is in excess of the bias one obtains from the CLF
approach (c.f. discussion surrounding \Fig{fig:LFbiasComp}). 

The above results are for a particular choice of $\Vs$ and in
principle, for a sufficiently large survey, $\Vs\sigma(\Vs)\rightarrow
0$, we are left with just the halo occupancy covariance and the
Poisson noise. Figure~\ref{fig:relerr} presents the ratio of the halo
occupancy variance with respect to the Poisson noise,
i.e. $\sigma^2[\phi(L_{\mu})]_{\rm O.V.}/ \sigma^2[\phi(L_{\mu})]_{\rm
  P.V.}=\widetilde{\Sigma}_{\mu\mu}$. This demonstrates that for the
brightest galaxies in a survey, the counts are dominated by Poisson
errors. However, for the fainter galaxies $L\lesssim L_{*}$ the halo
occupancy variance is roughly between $\sim0.6$--$0.8$ times the
Passion noise, independent of the survey volume. Note that this
fractional relation between the halo occupancy variance and the
Poisson errors holds exactly for both volume and flux-limited surveys.


\subsubsection{Correlation matrix}

\begin{figure}
\centerline{
  \includegraphics[width=8.5cm]{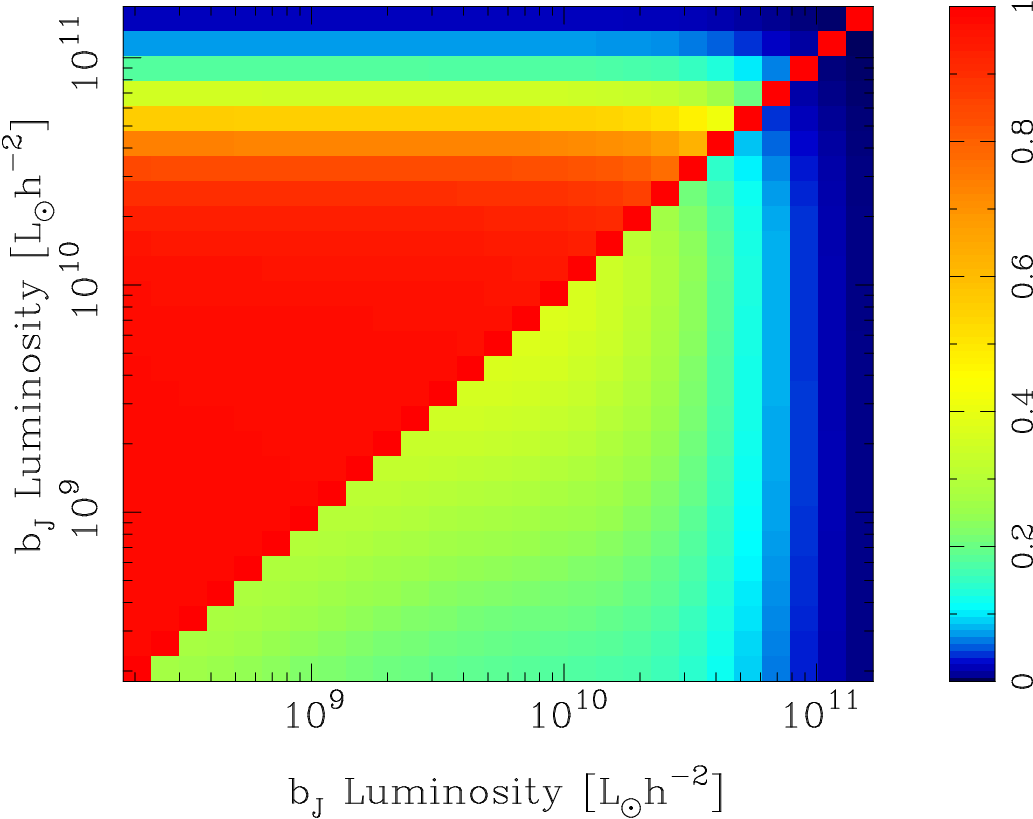}}
\caption{\small{Relative contributions of the sample and halo
    occupancy covariance to the correlation matrix, for a volume
    limited sample of galaxies. The upper left triangle represents the
    sample variance plus Poisson noise contribution, with the halo
    occupancy covariance set to zero.  The lower right triangle
    represents the halo occupancy covariance plus Passion noise
    contribution, with the sample variance set to zero.  }
\label{fig:LFCorrVolLimBreak}}
\end{figure}

Figure~\ref{fig:LFCorrVolLimBreak} presents the relative contributions
to the correlation matrix. The top left triangle shows the sample
covariance plus Poisson noise correlation matrix, with the halo
occupancy covariance set to zero. The bottom right triangle shows the
halo occupancy covariance plus Poisson noise, with the sample variance
set to zero. This demonstrates that, for the case of a volume limited
2dFGRS-like survey, with volume of size $V=0.125\Gpccube$, the
off-diagonal elements of the covariance matrix are entirely dominated
by the sample variance term. However, following our earlier discussion
from \S\ref{ssec:corrFL}, we now point out that if our survey was
sufficiently large, such that $\Vs\sigma(\Vs)\rightarrow0$, then the
matrix would still be correlated and that this would be given exactly
by the bottom right panel of Fig.~\ref{fig:LFCorrVolLimBreak}. The
figure shows that the minimum correlation coefficient that could be
obtained for galaxies with $L\lesssim L_{*}$ is roughly
\mbox{$r\sim0.2$--$0.4$}.

\begin{figure*}
\centerline{
  \includegraphics[width=8.5cm]{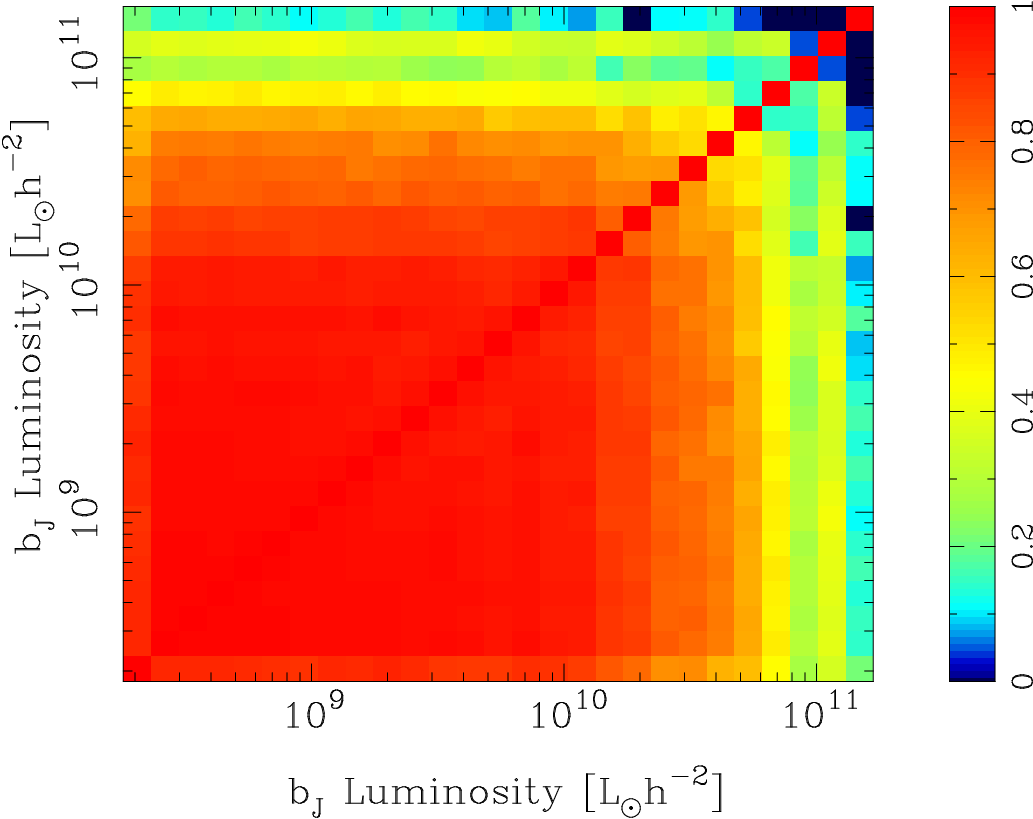}\hspace{0.3cm}
  \includegraphics[width=8.5cm]{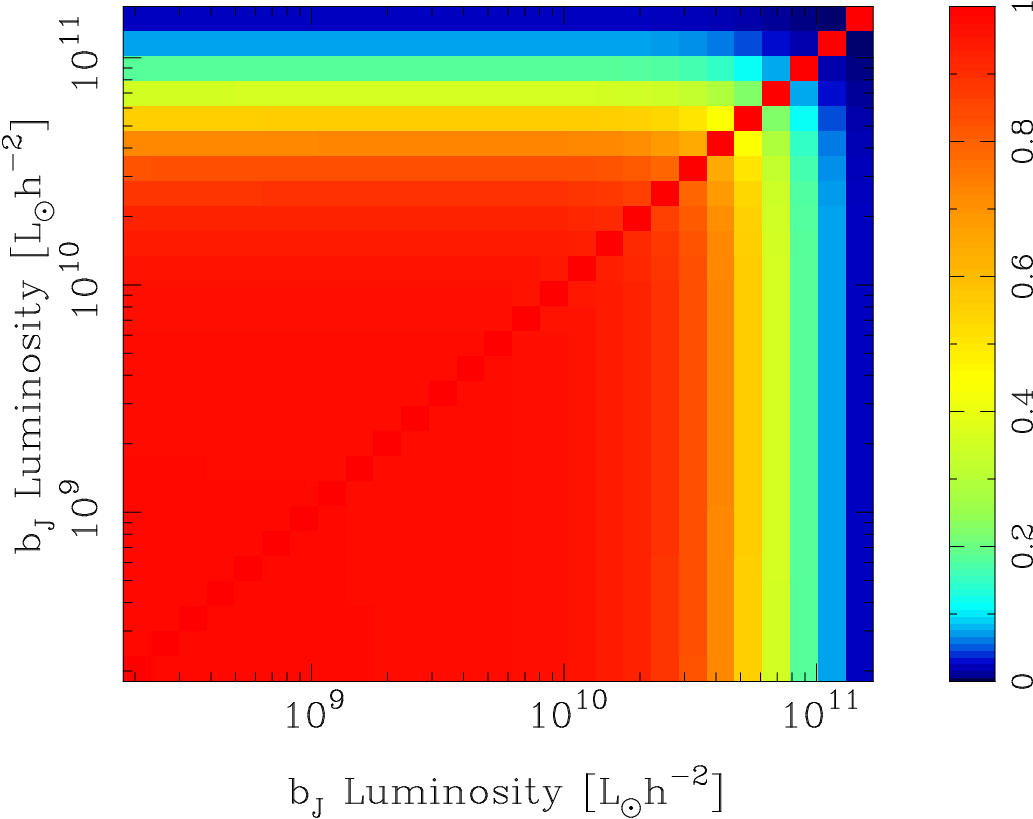}}
\caption{\small{Correlation matrix of luminosity function estimates
    for a volume limited sample of galaxies. {\em Left panel:} results
    obtained from the semi-analytic galaxies in the Millennium
    simulation. {\em Right panel:} results obtained from the
    theoretical model described in \Eqn{eq:crosscorr}. }
\label{fig:LFCorrVolLim}}
\end{figure*}

The left panel of Fig.~\ref{fig:LFCorrVolLim} presents the correlation
matrix constructed from the estimates of the covariance matrix
obtained through application of \Eqn{eq:EstCorrVolLim} to the SAM
data. The results show that the GLF estimates for galaxies with
luminosities $L<L_{*}$ are almost perfectly correlated,
i.e. $r\sim1$. This is a somewhat startling result, as it means that
if there is an upward fluctuation of one bin with respect to the mean
then all other bins share that same upwards fluctuation. As we will
discuss later, this has broad implications for how one fits models to
the measured GLF data.

The right panel of Fig.~\ref{fig:LFCorrVolLim}, presents our
theoretical predictions for the correlation matrix, evaluated using
\Eqn{eq:crosscorr} and the CLF model of \citet{Yangetal2003}.  We find
that the theoretical predictions are in remarkably good agreement with
the estimates from the SAM. The theoretical predictions are slightly
more correlated than the measurements from the SAM. This might be
attributed to the mis-match in the luminosity dependence of the galaxy
bias from the SAM and the 2dFGRS CLF model.


\subsection{Results: Flux limited samples}

Having validated our theoretical model for the covariance matrix of
the GLF, we now turn to the slightly more complicated case of
predicting the covariance matrix for a flux limited survey.

As described in \S\ref{sec:LFFluxLim} we must take into account that,
for a flux-limited survey, the observed volume depends on the
luminosity of the objects in question and the flux limit. We shall
take our fiducial survey to have an angular area coverage of roughly
$\Omega_{\rm s}\sim 1000\, {\rm deg^{2}}$. The survey volume will thus
be $\Vmu=\Omega_{\rm s}\chi^{\rm max}(L_{\mu})/3$, where $\chi^{\rm
  max}(L_{\mu})$ is given by \Eqn{eq:chimax}. In order to make
predictions we also need to know, $\sigma^2(L_{\mu},L_{\nu})$, and in
order to avoid dealing with the real complex survey geometry, we shall
make the approximation that the cone volume can be interpreted simply
as full-sky survey with radial dimension $R^{\rm
  max}(L_{\mu})\equiv[\Vmu]^{1/3}$. Hence we employ the window
function appropriate for a spherical-top-hat transformed into Fourier
space:
\be 
W(k|L_{\mu}) = \frac{3}{y^3}\left[\sin y-y\cos y\right] \ ; 
\ y\equiv kR^{\rm max}(L_{\mu})\ .
\ee
Further, we shall take the flux limit to be that equivalent to the
2dFGRS: $b_{J}=19.5$. In evaluating \Eqn{eq:chimax} we require the
conversion from $b_J$--luminosity to $b_J$--absolute magnitude, and we
do that using:
\be 
M(L_{\mu})=M_{\odot,b_{J}}-\frac{5}{2} 
\log_{10} \left[\frac{L_{\mu}}{L_{\odot}}\right] \ , \label{eq:MagLum}
\ee
where we have adopted $M_{\odot,b_{J}}=5.3$. Thus for galaxies with
the characteristic luminosity of the 2dFGRS, we have
$L_{*}=9.64\times10^9\Lsol$, which corresponds to
$M_{b_J}^*-5\log_{10}h=-19.66$, and the maximum distance out to which
they may be observed corresponds to $\chi_{\rm max}\approx680\Mpc$,
and with the volume being $\Vmu\approx0.13\Gpccube$.


\subsubsection{Diagonal errors}

\begin{figure}
\centerline{
\includegraphics[width=8.2cm]{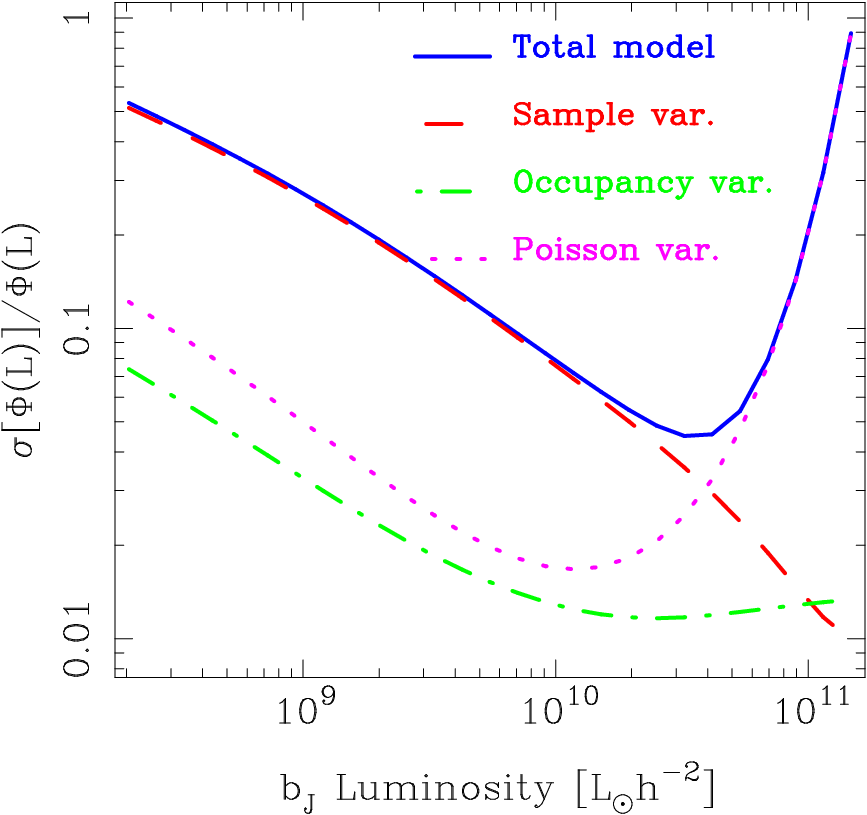}}
\caption{\small{Fractional errors on the galaxy luminosity function
    for a survey of angular size $\Omega_{\rm s}\sim 1000 \, {\rm
      deg}^2$, with limiting magnitude $b_{J}=19.5$ as a function of
    galaxy luminosity. The solid blue line presents the total
    prediction of the theoretical model given by
    \Eqn{eq:CovFluxLim}. The red dashed line denotes the contribution
    to the error from the sample variance; the green dot-dashed line
    corresponds to the error coming from the halo occupancy
    covariance; the magenta dotted line corresponds to the error from
    Poisson noise.}
\label{fig:LFDiagErrFluxLim}}
\end{figure}

Figure~\ref{fig:LFDiagErrFluxLim} presents the predictions for the
fractional errors on our fiducial survey. Considering again
\Eqn{eq:CovFluxLim}, we see that the fractional errors can be written:
\be 
\left.\frac{\sigma^2[\phi(L_{\mu})]}{\phi^2(L_{\mu})}\right|^{\rm FL}
= [b^{\rm g}_{\mu}]^2\sigma^{2}(L_{\mu},L_{\mu})+
\frac{1}{N^{\rm g}_{\mu}}+\frac{\Sigma^{\rm FL}_{\mu\mu}}{\phi_{\mu}^2} \ .
\ee
As for the case of the volume limited survey, the fractional errors
have three contributions: the sample variance, the Poisson noise and
the halo occupancy covariance. These three terms may also be described
by \Eqnss{eq:SV}{eq:OV}.

In the figure we see that as in the case of the volume limited sample
the fractional errors for the brightest galaxies are well described by
the Poisson error term. However, for galaxies at the characteristic
luminosity of the survey, the errors become dominated by the sample
variance term. The interesting change from the volume limited survey
is that when we consider the lower luminosity bins, we see that whilst
the sample variance term is still dominant, the contributions from the
Poisson variance and the halo occupancy variance are also significant.
This owes to the fact that $\Vmu$ is significantly smaller for
galaxies with $L\sim L_{*}$ than for the case of the volume limited
sample. Hence this leads to an increase in the Poisson shot noise for
these bins\footnote{Note that we are rescaling our errors to a volume
  of a given fiducial size, and for our fiducial flux-limited survey
  the effective volume is reduced for all galaxies fainter than the
  brightest luminosity bin that we employ.}. Finally, we recall that
owing to the fact that $\widetilde{\Sigma}_{\mu\mu}^{\rm
  FL}=\widetilde{\Sigma}_{\mu\mu}$, the relative strength of the halo
occupancy covariance to the Poisson noise is once more given by
\Fig{fig:relerr}.


\begin{figure}
\centerline{
  \includegraphics[width=8.5cm]{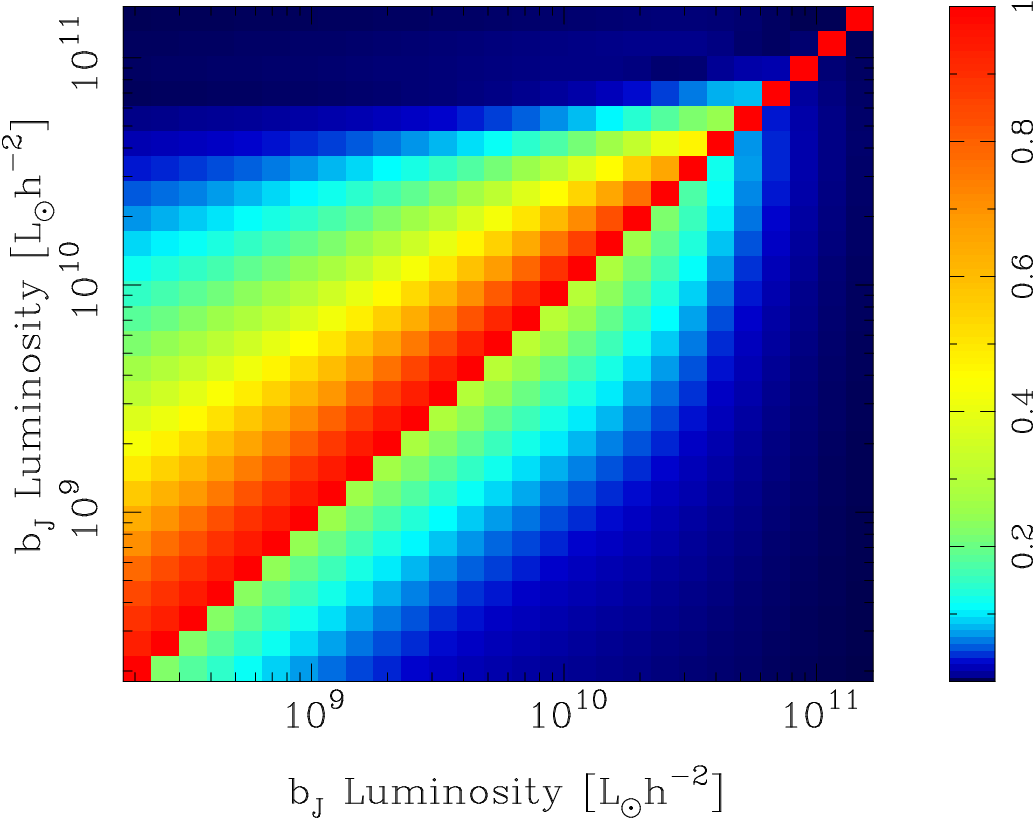}}
\centerline{
  \includegraphics[width=8.5cm]{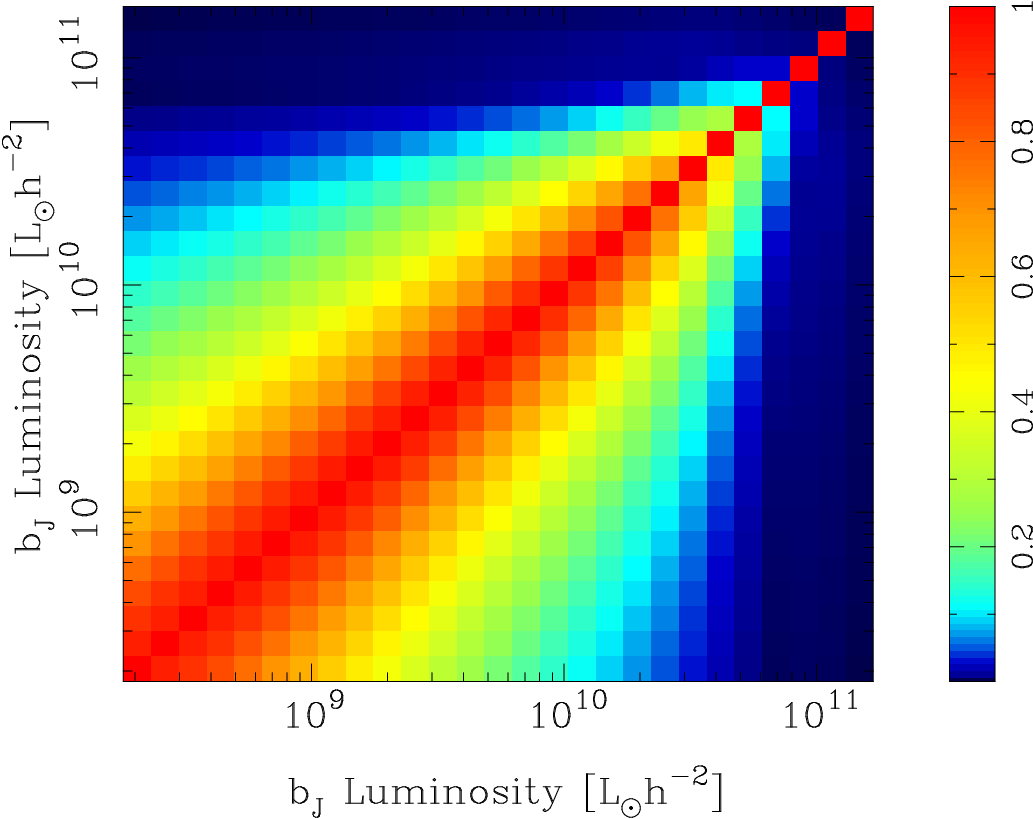}}
\caption{\small{Correlation matrix of luminosity function estimates
    for a survey of angular size $\Omega_{\rm s}\sim 1000 \, {\rm
      deg}^2$ and with a flux limit $b_{J}=19.5$. {\em Top panel:}
    Relative contributions of the sample and halo occupancy covariance
    to the correlation matrix. The upper left triangle represents the
    sample variance plus Poisson noise contribution, with the halo
    occupancy set to zero. The lower right triangle shows the same for
    the halo occupancy plus Poisson noise covariance contribution,
    but this time with the sample variance set to zero. {\em Bottom
      panel:} total correlation matrix as given by
    \Eqn{eq:crosscorrFluxLim}.}
\label{fig:LFCorrFluxLimBreak}}
\end{figure}


\subsubsection{Correlation matrix}

The top panel of Figure~\ref{fig:LFCorrFluxLimBreak} presents the
relative theoretical predictions for the correlation matrix. The top
left triangle shows the contributions to the correlation matrix that
come from the sample covariance plus Poisson noise, with the halo
occupancy covariance set to zero. The lower right corner shows the
same, but this time for the halo occupancy covariance plus Poisson
noise, with the sample variance set to zero.

The bottom panel of Fig.~\ref{fig:LFCorrFluxLimBreak} shows the
theoretical predictions for the total correlation matrix of GLF
estimates, as given by \Eqn{eq:crosscorrFluxLim}. We see that the
correlation matrix is almost diagonal for galaxies with $L>6L_{*}$.
However for galaxies with lower luminosities, the matrix becomes
strongly correlated. The correlations are not as strong as for the
Volume limited survey, however $r\sim1$ for luminosity bins that are
relatively close to one another. 

Clearly, for our fiducial survey, the correlation matrix is dominated
by the sample covariance, with a relatively small fraction of the
off-diagonal elements coming from the occupancy covariance. However,
for the case of a sufficiently large survey with
$\sqrt{\Vmu\Vnu}\sigma^2(L_{\mu},L_{\nu})\rightarrow0$, then the
off-diagonal elements of the correlation matrix do not vanish, but are
given by the bottom right triangle of \Fig{fig:LFCorrFluxLimBreak} top
panel.

  
\section{Impact on Parameter estimation}\label{sec:parameters}

We now explore the importance of including the covariance matrix when
estimating GLF parameters from observations.


\subsection{Methodology}

Suppose we have estimated the GLF from our survey using either E1 or
E2, depending on whether we have a flux or volume limited sample. We
now wish to interpret these estimates in terms of some model.  To do
this, let us adopt the Bayesian framework. The probability of
obtaining a data vector ${\bf x}$, given our model $\MM$ with
parameters ${\bm \theta}$, is described by the likelihood function
${\mathcal L}({\bf x}|{\bm \theta},\MM)$. A good choice for ${\mathcal
  L}$ is a multivariate Gaussian:
\be {\mathcal L}({\bf x}|{\bm \theta},\MM) = \frac{1}{(2\pi)^{N/2} \sqrt{ |\CC|}}
\exp\left[-\frac{1}{2}(\bx-{\bm \mu})^{T}\CC^{-1}
(\bx-{\bm \mu}) \right] \label{eq:likelihood}\ee
where $\bm\mu\equiv\bm\mu(\bm\theta)$ and ${\CC}\equiv
{\CC}(\bm\theta)$ are the model mean and model data covariance matrix,
both of which depend on the parameters ${\bm\theta}$; and $|\CC|$ is
the determinant of the matrix.  Using Bayes theorem, the likelihood is
directly related to the {\em posterior} probability distribution:
\be p(\bm\theta|{\bf x},\MM) = 
\frac{\Pi(\bm\theta|\MM){\mathcal L}(\bx|\bm\theta,\MM)}{p(\bx|\MM)} \ .
\ee
where $\Pi(\bm\theta|\MM)$ are a set of model {\em priors}, and
$p(\bx|\MM)$ is termed the {\em evidence}, which simply can be written
as a normalisation criterion: $p(\bx|\MM)=\int d {\bm\theta}
\Pi({\bm\theta}|\MM) {\mathcal L}(\bx|{\bm\theta},\MM)$.  The errors
on the model parameters may be obtained through the exploration of the
posterior distribution in the usual way
\citep{Pressetal1992,LewisBridle2002,Heavens2009}. Different models
$\MM_1$ and $\MM_2$ may then be compared using Bayesian model
selection methods \citep{Heavens2009}.

If the priors $\Pi(\bm\theta)$ are flat, then the posterior
$p(\bm\theta|{\bf x})$ is simply proportional to the likelihood
${\mathcal L}$.  Close to its maximum, at $\bm\theta_0$, we may Taylor
expand the logarithm of the posterior, and for flat priors the log
likelihood, to obtain:
\be \ln p(\bm\theta|\bx)
 \propto \ln {\mathcal L}({\bf x}|\bm\theta_0) - 
\frac{1}{2}
\sum_{\alpha,\beta}{\mathcal H}_{\alpha\beta}(\bm\theta_0)
\Delta\theta_{\alpha}\Delta\theta_{\beta} +\dots \ ,
\ee
where in the above
$\Delta\theta_{\alpha}\equiv\left(\theta_{\alpha}-\theta_{\alpha,0}\right)$
are deviations of the parameters from the fiducial values, \mbox{${\mathcal
  H}_{\alpha\beta}\equiv- \partial^2 \ln {\mathcal L} /
\partial\theta_{\alpha}\partial\theta_{\beta}$} is the Hessian matrix, and 
the first derivative vanished at the maximum. We may rewrite the above
expression for the posterior as,
\be p(\bm\theta|\bx) \approx 
\frac{\Pi(\bm\theta)}{p(\bx)} {\mathcal L}(\bm\theta_0)\exp\!\left[-
\frac{1}{2} \sum_{\alpha,\beta} \Delta\theta_\alpha 
{\mathcal H}_{\alpha\beta}(\bm\theta_0)
\Delta\theta_\beta \right]
\ .\ee
Thus ${\mathcal H}_{\alpha\beta}$ informs us about errors on the
parameters and how different parameters may be correlated with respect
to each other -- in the context of their effects on the data.

For the case of a multivariate Gaussian posterior, the marginalised
error on parameter $\theta_{\alpha}$, is given by
$\hat{\sigma}^2_{\alpha\alpha}=1/\left[ {\mathcal
    H}^{-1}\right]_{\alpha\alpha}$.  Since the likelihood itself
depends on the data, it is also a random variable. Taking an ensemble
average over many realizations of the data, we arrive at the Fisher
matrix:
\be {\mathcal F}_{\alpha\beta}=\left<{\mathcal
    H}_{\alpha\beta}\right>=-\left<\frac{\partial^2 \ln
    {\mathcal L}}{\partial\theta_\alpha\partial\theta_\beta}\right> \ .\ee
From the Fisher matrix one may obtain the expected marginalised error
on parameter $\theta_{\alpha}$ and the covariance between parameters
$(\theta_{\alpha},\theta_{\beta})$:
\be 
\sigma_{\alpha\alpha}\ge\sqrt{\left[ {\mathcal F}^{-1}\right]_{\alpha\alpha}}\ ; \hspace{1cm}
\sigma_{\alpha\beta}\ge\sqrt{\left[ {\mathcal F}^{-1}\right]_{\alpha\beta}}\ .
\ee
For a derivation of these error bounds see \citet{Heavens2009}.


\begin{figure}
\centerline{
\includegraphics[width=8.5cm]{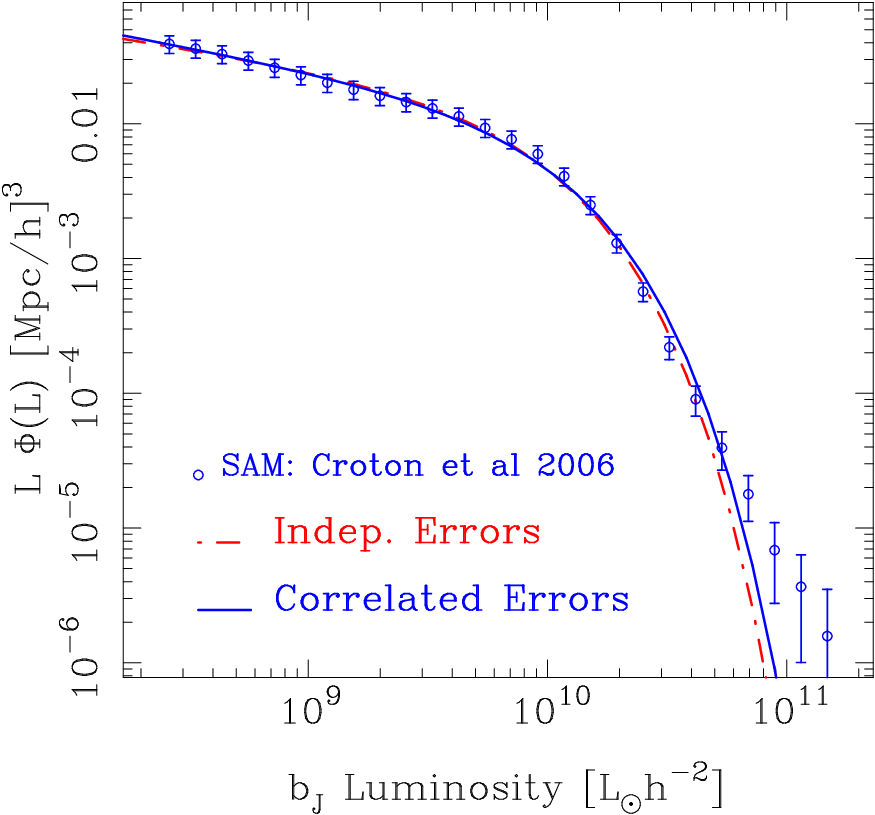}}
\caption{\small{Comparison of different Schechter function fits to
    semi-analytic model galaxy luminosity function data. The open
    points with errors denote the results from the
    \citet{Crotonetal2006} model data. The red dot-dashed line and
    solid blue lines correspond to the best-fit Schechter functions
    obtained when fitting using only the diagonal elements of the data
    covariance matrix and when using the full data covariance matrix,
    respectively.}\label{fig:bestfit}}
\end{figure}


Under the assumption that the likelihood is Gaussian in the data, c.f.
\Eqn{eq:likelihood}, then it can be shown that the Fisher matrix takes
on the special form \citep{Tegmarketal1997}:
\be {\mathcal F}_{\alpha\beta} =  \frac{1}{2} 
{\rm Tr}\left[\CC^{-1}\CC_{,\alpha}\CC^{-1}\CC_{,\beta}\right] +
{\bm\mu}_{\alpha}^{T}\CC^{-1}{\bm\mu}_{\beta} \ . \label{eq:FishGauss}
\ee
%


\begin{figure*}
\centerline{
\includegraphics[width=8.0cm]{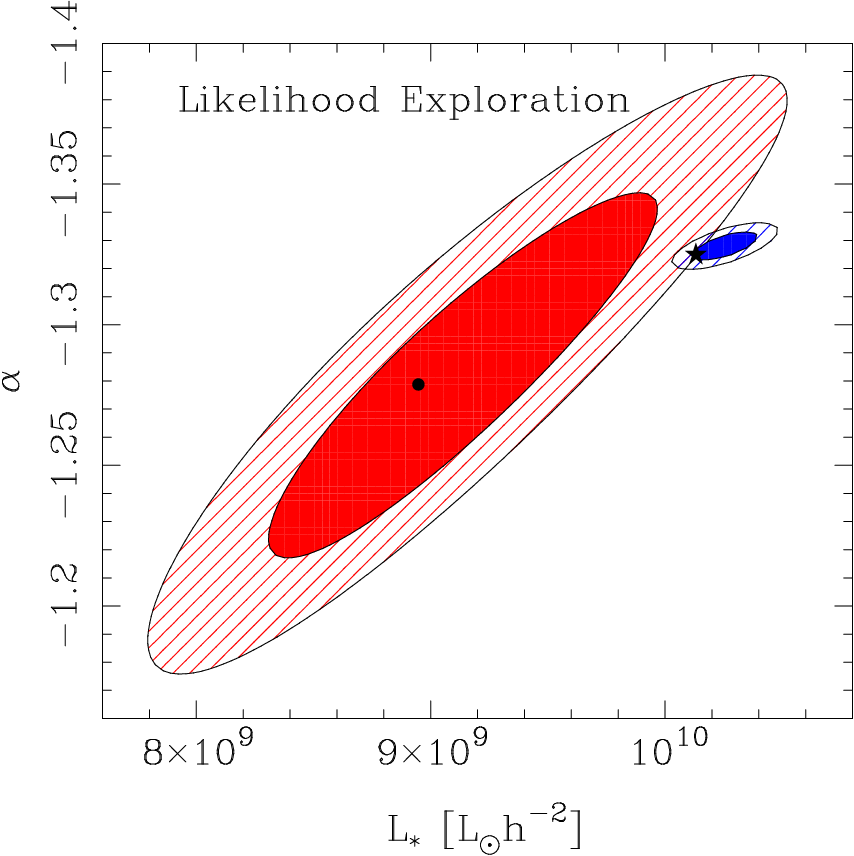}\hspace{0.3cm}
\includegraphics[width=8.0cm]{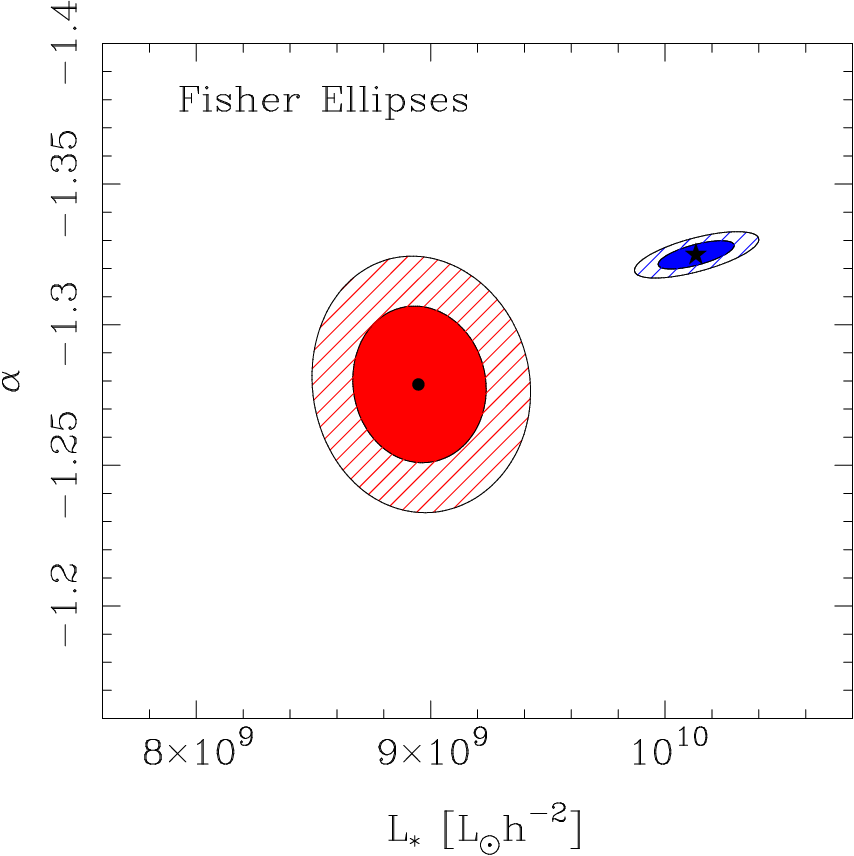}}
\caption{\small{2--D Likelihood contours for the luminosity function
    parameters $L_{*}$ and $\alpha$. {Left panel:} results obtained
    from a full exploration of the likelihood surface. The solid red
    ellipse and dashed red ellipse correspond to the 1-- and
    2--$\sigma$ confidence regions, respectively. These are obtained
    when we use only diagonal elements of the data covariance matrix
    in the parameter estimation. The solid blue ellipse and hatched
    blue ellipse, show the same, but when the full data covariance
    matrix is used. {\em Right panel:} Same as the left panel, except
    that the 1-- and 2--$\sigma$ confidence regions are obtained using
    the Fisher matrix formalism,
    c.f. \Eqn{eq:LFFish}.}\label{fig:likelihood}}
\end{figure*}


\subsection{Best-fit Schechter function for SAM data}

As a concrete example of our parameter estimation procedure, we now
find the best-fit Schechter function parameters that describe the SAM
data of \citet{Crotonetal2006}. We take the data for the volume
limited sample of SAM galaxies described in \S\ref{ssec:samLF}. Again,
we divide the full simulation volume into 64 equal sub-cubes and
estimate the GLF for each using estimator E1.  We then construct the
mean GLF and its covariance matrix, as described in
\S\ref{sec:results-sub1}. We shall estimate the best-fit parameters
for a survey region equivalent to a single sub-cube of size
$L=125\Mpc$.

We adopt a Schechter function GLF model, as described by
\Eqn{eq:Schechter}. As noted earlier, this has three parameters
\mbox{$\bm\theta=\{L_{*},\alpha,\phi_{*}\}$}. We treat $L_{*}$ and
$\alpha$ as free parameters and fix the normalisation $\phi_{*}$ by
the constraint that we desire to recover the mean number density of
galaxies in the volume that are above the luminosity cut $L_{\rm
  min}$:
\be 
\overline{n}_{\rm gal} = \int_{L_{\rm min}}^{\infty} dL \Phi(L|\bm\theta)  \ .
\ee
For the Schechter function this constraint is realised as:
\be \phi_{*} = \frac{\overline{n}_{\rm gal}}{\Gamma[\alpha+1,L_{\rm min}/L_{*}]} \ ,\ee
where $\Gamma[x,a]$ is the incomplete Gamma function. We then
construct the likelihood ${\mathcal L}$ as described by
\Eqn{eq:likelihood}. This function is maximised with respect to the
two free parameters, and we do this using an adaptive grid search
scheme. 

We find the 1-- and 2--$\sigma$ confidence regions of the likelihood
surface by identifying the contours in the ${\mathcal L}$--surface 
that satisfy:
\be p={\mathcal L}(\bx|\bm\theta_{0}) \exp\left[-\Delta\chi^2/2\right] \ ,\ee
where ${\mathcal L}(\bx|\bm\theta_{0})$ corresponds to the maximum of
the likelihood and $\Delta\chi^2=\{2.3,6.17\}$ for 1-- and 2--$\sigma$
contours, respectively.

The Fisher matrix approach of the previous section also provides us
with a means for estimating the covariance matrix of parameters.  From
\Eqn{eq:FishGauss}, and for a constant covariance matrix, the Fisher
matrix for the GLF is:
\be 
{\mathcal F}_{\alpha\beta} = \sum_{\mu,\nu} 
\frac{\partial \left[L\Phi(L_{\mu}|\bm\theta)\right]}{\partial \theta_{\alpha}}
\CC^{-1}_{\mu\nu}
\frac{\partial \left[L\Phi(L_{\nu}|\bm\theta)\right]}{\partial \theta_{\beta}} \ .
\label{eq:LFFish}
\ee
For the Schechter function parameters, the derivatives of interest are:
\ba
\frac{\partial \log \left[L\Phi(L|\bm\theta)\right]}{\partial L_{*}}
& = &  
\left[\frac{L-(\alpha+1)L_*}{L_*^2}\right] \ ;\\
\frac{\partial \log \left[L\Phi(L|\bm\theta)\right]}{\partial \alpha}
& = & 
\log\left[\frac{L}{L_*}\right]\ ; \\
\frac{\partial \log \left[L\Phi(L|\bm\theta)\right]}{\partial \phi_{*}}
& = & \frac{1}{\phi_{*}} \ .
\ea

Figure~\ref{fig:bestfit} shows the best-fit Schechter function
obtained when we fit the SAM GLF data using the full data covariance
matrix (solid blue line), which correctly takes into account the
effects of bin-to-bin correlations generated by the large-scale
structure in the volume. The figure also shows the best-fit Schechter
function model, obtained when we use only the diagonal elements of the
data covariance matrix for the parameter estimation (red dot dashed
line). It can be clearly seen that when we only use the diagonal
elements of the covariance matrix, the model is biased. This owes to
the fact that the fit gives more importance to the lower luminosity
bins, for which the errors are significantly smaller than for the
brighter bins. We may see this bias more clearly by exploring the
likelihood surface directly.

In the left panel of Figure~\ref{fig:likelihood} we show the 2-D
likelihood surfaces for the fitted parameters $L_*$ and $\alpha$, and
the 1-- and 2--$\sigma$ confidence limits.  The left panel shows the
results obtained when we employ a full exploration of the likelihood
surface. The blue ellipses show the results obtained when using the
full data covariance, and the red ellipses show the results when using
the diagonal elements of the covariance only. This figure shows the
best-fit values for $\{L_*,\alpha\}$ are only just consistent at the
2--$\sigma$ level. The best fit parameters are:
\be 
\bm\theta^{\rm full\, cov.}=\left\{
\begin{array}{rl}
L_* & = 1.023 \times 10^{10} \,[\Lsol]\\
\alpha & = 1.325 \\
\phi_{*} & =0.0122 \ [h^3{\rm Mpc}^{-3}]
\end{array}\right. \ .
\ee
\be 
\bm\theta^{\rm diag.\, cov.}=\left\{
\begin{array}{rl}
L_* & = 8.913 \times 10^{9} \,[\Lsol]\\
\alpha & =1.28 \\
\phi_{*} & = 0.0144\ [h^3{\rm Mpc}^{-3}]
\end{array}\right. \ .
\ee

In the right panel of Figure~\ref{fig:likelihood} we show the results
obtained when we employ the Fisher matrix formalism to calculate the
parameter covariance matrix. Considering the case where we used the
full data covariance matrix in the parameter estimation, we see that
the Fisher matrix predictions are in excellent agreement with the
full likelihood exploration. However, for the case where we used only
the diagonal elements of the covariance matrix in the fitting, we find
that the Fisher matrix errors are only qualitatively consistent with
the results for the full likelihood exploration. 


\section{Conclusions}\label{sec:conclusions}

In this paper we have investigated the galaxy luminosity function
(GLF) and what determines its error properties for various commonly
used estimators.

In \S\ref{sec:estimators} we described several commonly used
estimators for the GLF. We showed that Turner's estimator
\citep{Turner1979}, which attempts to correct for the effects of
large-scale structure on the GLF, is actually a biased estimator.

In \S\ref{sec:LFVolLim} we then focused on the simpler estimator of
\citet{Schmidt1968} for volume limited samples. Using a cluster
expansion approach, we showed that this estimator in the ensemble
limit was unbiased. We derived the covariance matrix of this estimator
and found that it was comprised of three terms. The first term takes
account of the sample variance, which depends on the biases of the
galaxies in the different luminosity bins and also the variance of
matter fluctuations in the survey volume. The second term was a simple
Poisson noise contribution. The third term was dubbed halo occupancy
covariance, and it arose due to the fact that several galaxies may be
hosted by the same dark matter halo. We proved that the necessary
requirement for sample covariance to vanish is:
$\sigma^2\Vs\rightarrow0$.

In \S\ref{sec:LFFluxLim} we investigated the $1/V^{\rm max}$ estimator
of \citep{Schmidt1968} for flux-limited surveys. We showed, in the
ensemble limit, that this is also an unbiased estimator. We derived
the covariance properties of the estimator. Similar to the case of the
volume limited estimator, this matrix could also be decomposed into
three terms: sample, Poisson and occupancy covariance terms. For the
sample variance term, the major difference was that one must consider
the cross-volume variance, since two distinct luminosity bins trace
two different sample volumes.  Again the necessary condition for the
sample variance to be subdominant was attained when
$\sigma^2[L_{\mu},L_{\mu}]\Vmu\rightarrow0$.

In \S\ref{sec:models} we described the semi-analytic model (SAM)
galaxy catalogue of \citet{Crotonetal2006} that we used to test our
theoretical model. We also summarised the conditional luminosity
function (CLF) model of \citet{Yangetal2003}. We showed that for the
volume limited estimator, both the SAM and CLF models were able to
reproduce the 2dFGRS GLF. We then investigated the luminosity
dependence of the galaxy bias, and showed that in the SAM model the
correlation functions for different luminosity binned samples showed
complicated scale-dependence. For $r>3\Mpc$ the bias was reasonably
flat. We measured the large-scale relative bias and found that the
brightest luminosity bin, $L>3\times10^{10}\Lsol$, showed a 50\%
larger bias relative to $L_*$ galaxies. For galaxies with $L<L_*$ we
found that the SAM model predicted a much flatter luminosity
dependence of the bias than was measured in the 2dFGRS. The CLF model,
by {\em fiat}, reproduced the 2dFGRS data. These results were in
agreement with earlier work
\citep{Lietal2007,Kimetal2009,Guoetal2011}.
 
In \S\ref{sec:results} we used the SAM galaxies to examine the
fractional errors on the GLF estimates from the volume limited
samples. We found that for the bright galaxies the fractional errors
were much larger than for the fainter bins. For these, however the
fractional error became flat below $L_*$. The errors were not reduced
as the number of galaxies in the bin dramatically increased. These
results were in excellent agreement with our predictions from the
theoretical model. This plateau effect was explained by the sample
variance being the dominant source of error for these luminosity bins.

Again using the SAM galaxies we estimated the covariance matrix of the
estimates of the GLF. We found that for the Millennium simulation
volume, the cross-correlation coefficient was $r<0.5$ only for
galaxies with $L>5\times 10^{10}\Lsol$. For lower luminosity
galaxies $r>0.5$ and at the faint end the matrix was almost perfectly
correlated. We showed that the theoretical predictions from our
theoretical model was again in excellent agreement.

We then used our theoretical model to make predictions for how the
errors would change for a GLF estimated from a flux limited
survey. For the fractional errors, the main differences from the
volume limited sample, were that for the low luminosity bins the
errors increased with decreasing luminosity. This owed to the reduced
surveyed volume for these bins. However, the sample variance was still
dominant on these scales. Exploring the covariance matrix, we found
that in this case the matrix was less correlated than for the volume
limited sample. However, the matrix was still highly correlated for
the galaxies with $L<L_{*}$. Again the off-diagonal covariance was
attributed to the sample variance term.

In \S\ref{sec:parameters} we explored the importance of including the
full data covariance matrix when interpreting observations in terms of
a given model. We showed that if one neglects the bin-to-bin
covariances in the luminosity function, then parameter estimates will
be biased. When fitting Schechter functions to data, we found that the
most seriously affected was the characteristic luminosity, which was
systematically under-estimated by 10-20\%.


\section*{Acknowledgements}

RES thanks Shaun Cole, Cristiano Porciani and Laura Marian for
comments on an early draft. RES also thanks the referee Jon Loveday
for constructive comments, and Brandt Robertson for bringing to our
attention his interesting paper on estimating GLF constraints. RES
acknowledges support from a Marie Curie Reintegration Grant and an
award for Experienced Researchers from the Alexander von Humboldt
Foundation.  The Millennium Run simulation used in this paper was
carried out by the Virgo Supercomputing Consortium at the Computing
Centre of the Max-Planck Society in Garching. The semi-analytic galaxy
catalogue is publicly available at
http://www.mpa-garching.mpg.de/galform/agnpape



\bibliographystyle{mn2e}

\input{LumFunCov.2.bbl}

\appendix


\section{Cluster count statistics}\label{app:cluster}

In this appendix we calculate $\left<N^{\rm c}_{\alpha}\right>_{P,s}$
and $\left<N^{\rm c}_{\alpha}N^{\rm c}_{\beta}\right>_{P,s}$. These
derivations follow from \citet{HuKravtsov2003} and
\citet{SmithMarian2011}.

Consider some large cubical patch of the Universe, of volume $\Vs$,
and containing $N$ clusters that possess some distribution of masses.
Let us subdivide the cluster population into a set of $N_m$ mass
bins. Let the number of clusters in the $\alpha^{\rm th}$ mass bin be
denoted $N^{\rm c}_{\alpha}$. We shall assume that the probability
that the volume contains $N^{\rm c}_{\alpha}$ clusters in the mass bin
$\alpha$, is a Poisson process:
\be P(N^{\rm c}_{\alpha}|m_{\alpha})= \frac{m_{\alpha}^{N^{\rm c}_{\alpha}}
\exp(-m_{\alpha})}{N^{\rm c}_{\alpha}!} \ .
\label{eq:defPoisson222}
\ee
For any quantity $X$ that depends on the number of clusters, we denote
the average over the sampling distribution--the Poisson process in
this case--as $\left<X \right>_P$. Thus, the average of $N^{\rm
  c}_{\alpha}$ over the sampling distribution can be written:
\be
\left<N^{\rm c}_{\alpha}\right>_P = m_{\alpha} 
\equiv \overline{m}_{\alpha} \left[1+\overline{b}_{\alpha} {\delta}_V(\bx) \right], 
\ee
where \mbox{$\overline{m}_{\alpha} = \overline{n}_{\alpha} V$} is the
expected number of counts averaged over the Poisson sampling
distribution and the density fluctuations ${\delta}_V(\bx)$ in the
volume. The volume of the survey and the volume-averaged overdensity
field, are written:
\ba 
\Vs & = & \int \dx' \, W(\bx'|\Vs)\ \ ;\\
{\delta}_V(\bx) & = & \frac{1}{\Vs}\int \dx' \, W(\bx'|\Vs) \delta(\bx) \ .
\ea
where $W(\bx|\Vs)$ is the window function for the survey and
$\overline{n}_{\alpha}$ and $\overline{b}_{\alpha}$ are given by
\Eqns{eq:nbarh}{eq:def_bias1}.
%
%

Following \citet{LimaHu2004} we take the likelihood of drawing a
particular set of cluster counts in the mass bins to be 
\mbox{${\bf N}\in\{N^{\rm c}_{1},\dots,N^{\rm c}_{N_m}\}$} in
the cells to be:
\be 
{\mathcal L}({\bf N}|\overline{\bf m},{\bf S}) = \int d^{\mathcal{N}}\!m
\left[\prod_{\alpha=1}^{N_m} P(N_{\alpha}|m_{\alpha}) \right] 
G({\bf m}|\overline{\bf m},{\bf S}) 
\label{eq:likelihood1} \ 
\ee
where \mbox{$\overline{\bf
    m}\in\{\overline{m}_{1},\dots,\overline{m}_{N_\alpha}\}$} is a
model for the counts in the cells, ${\mathcal N}=N_{m}$ and where it
was assumed that the statistics of the volume-averaged density field
are described by a multivariate Gaussian:
\be G({\bf m}|\overline{\bf
  m},{\bf S}) \equiv \frac{(2\pi)^{-N/2}}{|S|^{1/2}} \exp\left[-\frac{1}{2}
({\bf m}-\overline{\bf m})^{T}{\bf S}^{-1}({\bf m}-\overline{\bf m})\right] 
\ , 
\label{eq:likelihoodG}
\ee
where ${\bf S}$ is defined to be 
\be 
S_{\alpha\beta}\equiv 
\left< 
\left(m_{\alpha}-\overline{m}_{\alpha}\right)
\left(m_{\beta}-\overline{m}_{\beta}\right)
\right>_s 
\ee
Note, we refer to averages over the density field as sample averages
and for a quantity $X$, they will be denoted $\left<X\right>_s$.

At this point we may be more precise about what we mean by ensemble
and Poisson averages:
\be \left<X({\bf N})\right>_{P,s} \equiv \sum_{N^{\rm
    c}_{1}=0}^{\infty} \dots \sum_{N^{\rm c}_{N_m}=0}^{\infty}
{\mathcal L}({\bf N}|\overline{\bf m},{\bf S}) X({\bf N}) \ .
\label{eq:formal_average}
\ee
Equation~(\ref{eq:likelihood1}) can be simplified in two limits: If
$S_{\alpha}\ll \overline{m}_{\alpha}$, then the likelihood is the
product of Poisson processes; alternatively, in the limit of a large
number of counts in each cell, then the Poisson process becomes close
to Gaussian and the likelihood can be approximated as a Gaussian with
shifted mean and augmented covariance matrix:
\be {\mathcal L}({\bf N}|\overline{\bf m},{\bf S}) \approx 
G({\bf N}|\overline{\bf m},{\bf C}) \ \ ;\ \ {\bf C}=\overline{{\bf M}}+{\bf S}
\label{eq:likelihoodApprox}\ ,\ee
where $\overline{\bf
  M}\rightarrow\overline{M}_{\alpha\beta}=\delta^{K}_{\alpha,\beta}
\overline{m}_{\alpha}$. Note that in the above equation, the
approximate sign is used since negative number counts are formally
forbidden \citep[for a more detailed discussion of this
see][]{HuCohn2006}.

The covariance of the counts can be written
\ba 
\left<N^{\rm c}_{\alpha}N^{\rm c}_{\beta}\right>_{s,P} \!\!\! & = \!\!\! & \!\!\!
\sum_{N^{\rm c}_{1}=0}^{\infty}\dots \!\!\!\sum_{N^{\rm c}_{N_m}=0}^{\infty}
\!\!\!{\mathcal L}({\bf N}|\overline{\bf m},{\bf S}) 
N^{\rm c}_{\alpha}N^{\rm c}_{\beta} \nn\\
& = &  
\int d^{{\mathcal N}}\!m G({\bf m}|\overline{\bf m},{\bf S}) \nn \\
&  \times & \left[  m_{\alpha}m_{\beta}
 \tilde{\epsilon}_{\alpha\beta}
+ \left<\left(N^{\rm c}_{\alpha}\right)^2\right>_P\delta^{k}_{\alpha\beta}\right] \ .
\ea
where $\tilde{\epsilon}_{\alpha\beta}=1$ when $\alpha\ne\beta$ and 0
otherwise. Considering the second term on the right-hand-side of the
above equation, and recall that for the Poisson distribution we have:
\mbox{$\left<X^2\right>=\left<X\right>[1+\left<X\right>]$}. Hence, on
using this fact, and coupled with \mbox{$\overline{m}_{\alpha} =
  \overline{n}(M_{\alpha})\Delta M_{\alpha} V$} we find:
\ba 
\left<N^{\rm c}_{\alpha}N^{\rm c}_{\beta}\right>_{s,P}\!\!\! & = &
\!\!\!\int d^{{\mathcal N}}\!m G({\bf m}|\overline{\bf m},{\bf S}) \left[m_{\alpha}m_{\beta} +
m_{\alpha} \delta^{K}_{\alpha,\beta}\right]  \nn \\
&  = &  
\left[S_{\alpha\beta}+\overline{m}_{\alpha}\overline{m}_{\beta}
+\overline{m}_{\alpha} \delta^{K}_{\alpha,\beta}\right] \ .
\ea
%


\section{Incorporating magnitude errors}\label{sec:magerrs}

\def\obs{{\rm o}}


The above analysis has so far included errors induced in the GLF that
arise from large-scale structures and also the occupancy of galaxies
in haloes. We now examine how the above results are modified in the
presence of calibration errors in the magnitudes of the galaxies.
Again, we shall look to the results from the 2dFGRS for illustration.

We take account of the mapping between the true luminosity $L$ and the
observed $L^{\obs}$ in the following way: the observed GLF can be
written
\be 
\phi(L^{\obs}_{\mu})=\int_{L^{\obs}_{\mu}}^{L_{\mu+1}^{\obs}} 
\dL^{\obs} \int_0^{\infty} \dL p(L^{\obs}|L) \phi(L) \ ,
\ee
where galaxies are observed with luminosities in the bin
$L^{\obs}_{\mu}<L^{\obs}\le L^{\obs}_{\mu+1}$. In the above the key
new ingredient is the probability distribution for obtaining a
luminosity $L^{\obs}$ given the underlying true luminosity $L$. In
\citet{Norbergetal2002bshort}, the observed $b_J$-band magnitudes,
$m^{\obs}$, of the 2dFGRS galaxies were found to have a calibration
error that was well described by a Gaussian with width
$\sigma_m=0.15$, with underlying true mean magnitude $m$. Hence,
\ba p(L^{\obs}|L) & = & p_{\rm G}(m^{\obs}|m)
\left|\frac{dm^{\obs}}{\dL^{\obs}}\right| \nn \\ 
& & \hspace{-1.0cm}= \frac{1}{\sqrt{2\pi}\sigma_M}
\exp\left[-\frac{(m-m^{\obs})^2}{2\sigma_M^2}\right] 
\left|\frac{dm^{\obs}}{\dL^{\obs}}\right|
\nn \\
& & \hspace{-1.0cm}= -\frac{2L^{\obs} \log_{e}10}{5\sqrt{2\pi}\sigma_m} 
\exp\left[-\frac{25\left(\log_{10}L/L^{\obs}\right)^2}{8\sigma_M^2}\right] \ ,
\ea
where in the above equations we used the relation
$L/L^{\obs}=10^{-2/5(m-m^{\obs})}$, to compute the Jacobean of the
coordinate transformation: $|\dL^{\obs}/dm^{\obs}|=-2L^{\obs}
\log_{e}10/5$.

Thus, with magnitude error uncertainties included, the covariance
matrix becomes,
\ba 
{\mathcal C}[L^{\obs}_{\mu},L^{\obs}_{\nu}] & = &
\int_{L^{\obs}_{\mu}}^{L_{\mu}^{\obs}+1} 
 \dL_1^{\obs} 
\int_{L^{\obs}_{\nu}}^{L_{\nu+1}^{\obs}}
 \dL_2^{\obs}  
\nn \\
& & \hspace{-2.0cm}
\times  \int \dL_1 p(L_1^{\obs}|L_1) \int \dL_2 p(L_2^{\obs}|L_2) 
\, {\mathcal C}[L_{1},L_{2}] 
\ea
On inserting \Eqn{eq:CovVolLim} for the true covariance, the observed
covariance can be written
\ba 
& & \hspace{-0.5cm}
{\mathcal C}[L_{\mu}^{\obs},L_{\nu}^{\obs}]=
\widetilde{\phi b^{g}}(L^{\obs}_{\mu})\widetilde{\phi b^{g}}(L^{\obs}_{\nu})\sigma^2(\Vs)
+\frac{\widetilde{\phi}(L^{\obs}_{\mu})  \delta^{K}_{\mu,\nu}}{\Vs \Delta L^{\obs}_{\mu}} 
\nn \\
& & + \frac{1}{\Vs}
\int dM_1 n(M_1) 
\widetilde{\phi}(L^{\obs}_{\mu}|M_1) 
\widetilde{\phi}(L^{\obs}_{\nu}|M_1)\ .
\label{eq:covariance}\ea
where we have defined three new terms:
\ba 
\!\!\widetilde{\phi}(L^{\obs}_{\mu}) & \!\!\equiv & \!\!\!
\int_{L^{\obs}_{\mu}}^{L_{\mu+1}^{\obs}}  \frac{\dL_1^{\obs}}{\Delta L_{\mu}}
\int_0^{\infty} \dL_1 p(L_1^{\obs}|L_1) \phi(L_1)\ ;\\
\!\!\widetilde{\phi b^{g}}(L^{\obs}_{\mu}) &  \!\! \equiv &  \!\!\! 
\int_{L^{\obs}_{\mu}}^{L_{\mu+1}^{\obs}}\!  \frac{\dL_1^{\obs}}{\Delta L_{\mu}}
\int_0^{\infty} \! \dL_1 p(L_1^{\obs}|L_1) \phi(L_1)b^g(L_1)\ ;\\ 
\!\!\widetilde{\phi}(L^{\obs}_{\mu}|M) &  \!\! \equiv &  \!\!\! 
\int_{L^{\obs}_{\mu}}^{L_{\mu+1}^{\obs}}  \frac{\dL_1^{\obs}}{\Delta L_{\mu}}
\int_0^{\infty} \dL_1 p(L_1^{\obs}|L_1) \phi(L_1|M) \ .
\ea
In the limit where the luminosity bins are sufficiently narrow that
the integrand does not vary across the bin, then the first integral in
the above equations may be approximated by the central value of the
integrand, in accordance with the mean value theorem. Furthermore,
since we take the error in the magnitude distribution to be a Gaussian
of width $\sigma_m$, the limits of the second integral can be
restricted to be $L_{\rm max}(L^{\obs})$ and $L_{\rm
  min}(L^{\obs})$. Hence
\ba 
\!\!\widetilde{\phi}(L^{\obs}_{\mu}) & \!\!\approx & \!\!\!
\int_{L_{\rm min}(L^{\obs})}^{L_{\rm max}(L^{\obs})} \dL_1 p(L_1^{\obs}|L_1) \phi(L_1)\ ;\\
\!\!\widetilde{\phi b^{g}}(L^{\obs}_{\mu}) &  \!\! \approx &  \!\!\! 
\int_{L_{\rm min}(L^{\obs})}^{L_{\rm max}(L^{\obs})} \! \dL_1 p(L_1^{\obs}|L_1) \phi(L_1)b^g(L_1)\ ;\\ 
\!\!\widetilde{\phi}(L^{\obs}_{\mu}|M) &  \!\! \approx &  \!\!\! 
\int_{L_{\rm min}(L^{\obs})}^{L_{\rm max}(L^{\obs})} \dL_1 p(L_1^{\obs}|L_1) \phi(L_1|M) \ .
\ea
In practice, the upper and lower bounds on the integrals are computed
by allowing the minimum `true' magnitude, which contributes to an
observed magnitude bin, to be 4$\sigma$ away from the mean,
respectively.  This gives,
\ba
L_{\rm max}/ L^{\obs} & = & 10^{-2/5(m^{\rm min}-m^{\obs})} = 10^{8/5\sigma_m} \\
L_{\rm min}/ L^{\obs} & = & 10^{-2/5(m^{\rm max}-m^{\obs})} = 10^{-8/5\sigma_m}  \ .
\ea
On adopting the appropriate value for the 2dFGRS, $\sigma_m=0.15$,
this leads us to adopt the integral limits $L_{\rm max}=1.74 L^{\obs}$
and $L_{\rm min}=0.575 L^{\obs}$.


\end{document}

%% file: defs.tex
\newcommand{\be}{\begin{equation}}
\newcommand{\ee}{\end{equation}}
\newcommand{\ba}{\begin{eqnarray}}
\newcommand{\ea}{\end{eqnarray}}

\newcommand\Eqn[1]     {Eq.\,(\ref{#1})}
\newcommand\Eqns[2]    {Eqs\,(\ref{#1}) and~(\ref{#2})}
\newcommand\Eqnss[2]   {Eqs\,(\ref{#1})--(\ref{#2})}

\newcommand\Fig[1]     {Fig.\,{\ref{#1}}}

\newcommand\nn         {\nonumber}

\def\CC{{\rm \bf C}}
\def\MM{{\mathcal M}}

\def\pp1{{\prime}}
\def\pp2{{\prime\prime}}

\def\2D{{\rm 2D}}

\def\Vs{{V_{\rm s}}}
\def\Vmu{V_{\mu}^{\rm max}}
\def\Vnu{V_{\nu}^{\rm max}}

\def\g{{\rm g}}

\def\bx{{\bf x}}

\def\bk{{\bf k}}

\def\1Loop{{\rm 1Loop}}

\def\Msol{h^{-1}M_{\odot}}
\def\Lsol{h^{-2}L_{\odot}}
\def\kpc{\, h^{-1}{\rm kpc}}
\def\Mpc{\, h^{-1}{\rm Mpc}}

\def\Gpccube{\, h^{-3} \, {\rm Gpc}^3}
\def\kMpc{\, h \, {\rm Mpc}^{-1}}

\def\dx{{\rm d}^3{\!\bf x}}
\def\dk{{\rm d}^3{\bf k}}

\def\dL{{\rm d}L}

\def\nbar{\bar{n}}

\def\fun#1#2{\lower3.6pt\vbox{\baselineskip0pt\lineskip.9pt
        \ialign{$\mathsurround=0pt#1\hfill##\hfil$\crcr#2\crcr\sim\crcr}}}

\def\obs{{\rm obs}}

\def\mnras{{Mon.~ Not.~ R.~ Astron.~ Soc.~}}

\def\prd{{Phys.~ Rev.~ D.~}}

\def\apj{{Astrophys.~ J.~}}

\def\nat{{Nature (London)~}}

\def\mnras{{MNRAS}}

\def\prd{{PRD}}

\def\apj{{ApJ}}

\def\aap{{A\&A}}
\def\nat{{Nature}}

\def\physrep{{Phys.~ Rep.~}}

\def\araa{Anual Reviews of Astronomy and Astrophysics}
\def\aj{Astronomical Journal}